\documentclass[twocolumn,aps,superscriptaddress,showkeys,amsmath]{revtex4}		
\usepackage{graphicx}
\usepackage{color} 
\begin{document}  

\title{Phase-dependent fluctuations of resonance fluorescence near the 
coherent population trapping condition}  

\author{O. de los Santos-S\'anchez} 
\email{octavio.desantos@gmail.com} 
\affiliation{Escuela de Ingenier\'{\i}a y Ciencias, Instituto Tecnol\'ogico y de Estudios 
Superiores de Monterrey, \\  Avenida San Carlos 100, Campus Santa Fe, Ciudad 
de M\'exico, 01389, M\'exico}
\author{H. M. Castro-Beltr\'an} 
\email{hcastro@uaem.mx} 
\affiliation{Centro de Investigaci\'on en Ingenier\'ia y Ciencias Aplicadas, 
Instituto de Investigaci\'on en Ciencias B\'asicas y Aplicadas, Universidad 
Aut\'onoma del Estado de Morelos, Avenida Universidad 1001, 62209 
Cuernavaca, Morelos, M\'exico}

\date{\today}

\begin{abstract}
We study phase-dependent fluctuations of the resonance fluorescence of a single 
$\Lambda$-type three-level atom in the regime near coherent population trapping, 
i.e., alongside the two-photon detuning condition. To this end, we employ the method 
of conditional homodyne detection (CHD) which considers squeezing  in the weak 
driving regime, and extends to non-Gaussian fluctuations for saturating and strong 
fields. In this framework, and using estimated parameter settings of the resonance 
fluorescence of a single trapped $^{138} \mathrm{Ba}^{+}$ ion, the light scattered 
from the probe transitions is found to manifest a non-classical character and 
conspicuous asymmetric third-order fluctuations in the amplitude-intensity 
correlation of CHD. 
\end{abstract}

\maketitle

\section{Introduction}  
Quantum interference effects in the interaction between matter and light, 
epitomized by coherent population trapping (CPT) and electromagnetically 
induced transparency (EIT), have extensively been studied, both theoretically 
and experimentally, over the past decades \cite{Arimondo,FlIM05}. The most 
common level structure to enable these effects is the three-level system in the 
$\Lambda$ configuration ($\Lambda$-3LA). Although early research in this 
regard was primarily focused on ensembles of atomic constituents, 
state-of-the-art experimental developments in atomic spectroscopy have made 
it possible to realize EIT with a single atom in free space \cite{SHG+10}. 
Indeed, these achievements have paved the way for exploring new avenues of 
spectroscopic analyses, besides their potential applications in the thriving field 
of quantum information, demonstrating, for instance, the viability of 
single-atom-based optical logic gates and quantum memories \cite{HPL+09}.  

Both CPT and EIT are based on the cancellation of absorption when two lasers 
are detuned equally on adjacent transitions, thus stopping further fluorescence. 
Near this two-photon detuning condition, large quantum fluctuations are thus 
expected. Phase-sensitive fluctuations of the electromagnetic field, usually 
characterized by the phenomenon of squeezing, are of particular interest. 
Squeezing is the shrinking of a field's quadrature fluctuations at the expense of 
increasing those of its conjugate, and is signaled by negative spectra or variance 
below the shot noise level. For the resonance fluorescence of a single two-level 
atom, squeezing was first predicted almost forty years ago \cite{WaZo81,CoWZ84}, 
but it was only very recently that squeezing of a two-level quantum dot was 
observed \cite{SHJ+15}. This achievement required overcoming the large 
collection losses of resonance fluorescence and the quantum detection losses 
of the standard balanced homodyne detection (BHD) technique. These issues 
were addressed, respectively, by the higher photon collection geometry allowed 
by the quantum dot, and by using a method called homodyne correlation 
measurement (HCM)  \cite{Vogel91,Vogel95,KVM+17}. 

The HCM method realizes an intensity-intensity correlation of the light of a 
previously selected quadrature; by measuring for several phases of a weak local 
oscillator, the method gives access to the variance (squeezing) \cite{SHJ+15} and 
a third-order moment of the field. The latter signals the evolution of the field after 
a photon was detected, as was demonstrated for the resonance fluorescence of a 
$\Lambda$-3LA \cite{GRS+09}, in a driving regime not weak enough to obtain 
squeezing, and far from EIT. The third-order moment is a reachable step above 
squeezing in the quest for high-order non-classicality \cite{ScVo05,ScVo06}. 

Conditional homodyne detection (CHD) is another measurement scheme 
capable of detecting phase-dependent fluctuations with high efficiency owing 
to its conditional character \cite{CCFO00,FOCC00,CFO+04}. It consists of BHD 
on the cue of photons recorded in a separate photodetector, giving direct access 
to the third-order moment of the field. Squeezing is measured if the source is 
weakly excited (in fact, the first motivation for the scheme) since in this case the 
third-order fluctuations of the field are small. However, these fluctuations, 
non-negligible for stronger excitation, are no less interesting: CHD goes beyond 
squeezing \cite{hmcb10,CaGH15} and reveals the non-Gaussian character of a 
source. 

One manifestation of non-Gaussian fluctuations is the asymmetry of the field's 
amplitude-intensity correlation whenever two or more transitions compete \cite{DeCC02}; it is not observed in the resonance fluorescence of a two- or 
three-level atom driven by a single laser \cite{hmcb10,CaGH15,CaRG16}. While 
this asymmetry was readily observed for cavity QED systems both numerically \cite{CCFO00, DeCC02} and experimentally \cite{FOCC00}, it has been the 
resonance fluorescence of several 3LA systems that have provided clear 
theoretical access to the understanding of the asymmetry 
\cite{MaCa08,GCRH17,XGJM15,XuMo15,GaJM13,WaFO16}. 
More recent accounts of asymmetric correlations are found in plasmonics 
\cite{Santos19} and collective cavity QED \cite{Zhao+20}.

In the experiment outlined in \cite{GRS+09}, squeezing, far from the two-photon detuning, was explored in the weak field regime. In keeping with the same spirit, quantum fluctuations of the light scattered by a coherently driven V-type 3LA have thoroughly been analyzed \cite{GCRH17}. In this work, near the two-photon detuning, we investigate, within the framework of CHD, the adjoining effect of CPT on the phase-dependent quantum fluctuations of 
the emitted light of the probe transition of a $\Lambda$-3LA by amplitude-intensity 
correlations. We follow closely the experimental 
conditions of observation of EIT in single $^{138} \mathrm{Ba}^+$ resonance 
fluorescence of Ref.~\cite{SHG+10}, where saturation is present, and we find the 
fluctuations to be predominantly non-Gaussian.

Our work is structured by introducing the atom-laser model in Section 2, discussing 
the role of coherent population trapping on the state populations and on the 
emission spectrum; section 3 is devoted to the theory of conditional homodyne 
detection and the analysis of quadrature fluctuations via the associated 
amplitude-intensity correlation. We study, in section 4, the quadrature fluctuations 
in the spectral domain, including squeezing and variance. Finally, in section 5, we 
present our conclusions and an appendix shows additional calculations.   

\begin{figure}[t!]
\includegraphics[width=6.5cm, height=4.0cm]{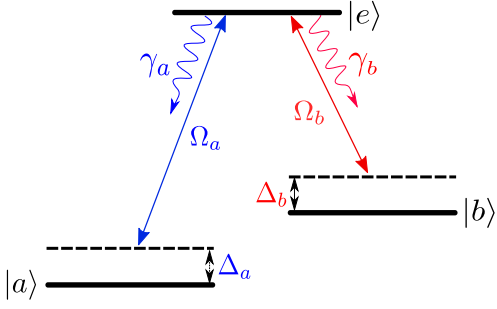} 
\caption{\label{fig:3LA} 
Scheme of the $\Lambda$ three-level atom with spontaneous decay rates  
$\gamma_a, \gamma_b$, interacting with lasers with Rabi frequencies 
$\Omega_a, \Omega_b$ with detunings $\Delta_a,\Delta_b$. }
\end{figure}
\section{Model} \label{sec:theory}
\subsection{Atom-Laser Interaction}
Our system, pictorially represented in Fig.~\ref{fig:3LA}, consists of a 
$\Lambda$-type three-level atom ($\Lambda$-3LA) with a single excited state 
$|e \rangle$ coupled by a monochromatic laser 
with Rabi frequency $\Omega_a$ to the ground state $|a \rangle$ and decay rate $\gamma_a$, and to a 
long-lived state $|b \rangle$ by a monochromatic laser with Rabi frequency 
$\Omega_b$ and decay rate $\gamma_b$. Decay from $|b \rangle $ to 
$|a \rangle$ is dipole-forbidden. Henceforth, the fields driving the 
$|a \rangle \to |e\rangle$ and $|b\rangle \to |e \rangle$ transitions will be referred 
to as the probe and control fields. We define the atomic operators as 
$\hat{\sigma}_{jk} = |j \rangle \langle k|$. 

Under the above considerations, the system's evolution, in free space, and in 
the frame rotating at the laser frequencies, $\nu_{a}$ and $\nu_{b}$, is 
governed by the master equation 
$\dot{\tilde{\rho}}(t)=-i [\hat{H}, \tilde{\rho}] 
+\sum_{j} \frac{\gamma_j}{2} \mathcal{L}_{\hat{\sigma}_{je}}[\tilde{\rho}]$, 
in which 
\begin{equation}
\hat{H} =  \sum_{j=a,b}  - \Delta_j \hat{\sigma}_{jj} 
	+ \frac{\Omega_j}{2} ( \hat{\sigma}_{ej} +\hat{\sigma}_{je})
\end{equation}
is the atom-laser Hamiltonian, and $\Delta_{j} = \omega_{ej}-\nu_{j} $ labels 
the individual atom-laser detunings. Dissipation is accounted for by the action 
of the Lindblad generator 
$\mathcal{L}_{\hat{O}}[\tilde{\rho}] = 2\hat{O} \tilde{\rho} \hat{O}^{\dagger} 
- \hat{O}^{\dagger} \hat{O} \tilde{\rho} -\tilde{\rho} \hat{O}^{\dagger} \hat{O} $, 
with $\hat{O}=\hat{\sigma}_{je}$. With the help of the relationship 
$\hat{\sigma}_{jk} \hat{\sigma}_{lm} = \hat{\sigma}_{jm} \delta_{kl}$, the master equation can be explicitly recast as 
\begin{equation}  	\label{masterEq} 
\dot{\tilde{\rho}}(t) = -i [\hat{H}, \tilde{\rho}] 
	+\sum_{j=a,b} \gamma_j \tilde{\rho}_{ee} \hat{\sigma}_{jj} 
	- \frac{\gamma_j}{2} \left( \hat{\sigma}_{ee} \tilde{\rho} 
	+\tilde{\rho} \hat{\sigma}_{ee} \right).
\end{equation}
With the relation $\langle \hat{\sigma}_{jk} \rangle = \tilde{\rho}_{kj}$, 
Eq.~(\ref{masterEq}) allows us to arrive at the following set of linear 
equations for populations: 
{\setlength\arraycolsep{2pt}
\begin{eqnarray}
 \langle \dot{\hat{\sigma}}_{aa} \rangle 
 &=& - i \frac{\Omega_a}{2} (\langle \hat{\sigma}_{ae} \rangle 
 	- \langle \hat{\sigma}_{ea} \rangle)
 	+\gamma_a \langle \hat{\sigma}_{ee} \rangle, 	\label{eq:pop1} \\
\langle \dot{\hat{\sigma}}_{bb} \rangle 
&=& - i \frac{\Omega_b}{2} (\langle \hat{\sigma}_{be} \rangle 
	- \langle \hat{\sigma}_{eb} \rangle)
 	+\gamma_b \langle \hat{\sigma}_{ee} \rangle, 	\label{eq:pop2} \\
\langle \dot{\hat{\sigma}}_{ee} \rangle 
&=& i \frac{\Omega_a}{2} (\langle \hat{\sigma}_{ae} \rangle 
		- \langle \hat{\sigma}_{ea} \rangle)  
  + i \frac{\Omega_b}{2} (\langle \hat{\sigma}_{be} \rangle 
		- \langle \hat{\sigma}_{eb} \rangle) \nonumber \\  
	& & -  \gamma \langle \hat{\sigma}_{ee} \rangle, \label{eq:pop3}
\end{eqnarray}}
with $\gamma= \gamma_a +\gamma_b$, and coherences:
{\setlength\arraycolsep{2pt}
\begin{eqnarray}
\langle \dot{\hat{\sigma}}_{ab} \rangle &=& 
	 i\frac{ \Omega_a}{2} \langle \hat{\sigma}_{eb} \rangle 
  	- i \frac{\Omega_b}{2} \langle \hat{\sigma}_{ae} \rangle 
  	- i (\Delta_a -\Delta_b) \langle \hat{\sigma}_{ab} \rangle, \label{eq:coh1} \\
\langle \dot{\hat{\sigma}}_{ae} \rangle &=& 
	i \frac{\Omega_a}{2} (\langle \hat{\sigma}_{ee} \rangle 
	- \langle \hat{\sigma}_{aa} \rangle)
	-i\frac{\Omega_b}{2} \langle \hat{\sigma}_{ab} \rangle \nonumber \\ 
	&& -  \left( \frac{\gamma}{2} 
	+ i \Delta_a \right)  \langle \hat{\sigma}_{ae} \rangle, 	\label{eq:coh2} \\
\langle \dot{\hat{\sigma}}_{be} \rangle &=& 
 	-  \frac{i\Omega_a}{2}  \langle \hat{\sigma}_{ba} \rangle 
	+  \frac{i\Omega_b}{2} (\langle \hat{\sigma}_{ee} \rangle 
		- \langle \hat{\sigma}_{bb} \rangle) \nonumber \\ 
	&& - \left( \frac{\gamma}{2} 
	+i \Delta_b \right) \langle \hat{\sigma}_{be} \rangle, \label{eq:coh3} \\
 \langle \dot{\hat{\sigma}}_{jk} \rangle 
	&=&  \langle \dot{\hat{\sigma}}_{kj} \rangle^{\ast}. \label{eq:coh4}
\end{eqnarray}}

The solution to these equations (a set of nine Bloch equations) is to be 
obtained numerically and their structure will facilitate the assessment of the 
sought correlation functions via the quantum regression formula, combined 
with  the employment of matrix methods. For later use, we define the values 
of the atomic operators in the steady state as 
\begin{equation}
\langle \hat{\sigma}_{jk} (t \to \infty) \rangle 
= \langle \hat{\sigma}_{jk} \rangle_{ss} = \alpha_{jk} .
\end{equation}
%
\begin{figure}[t!]
\includegraphics[width=8.7cm, height=5.5cm]{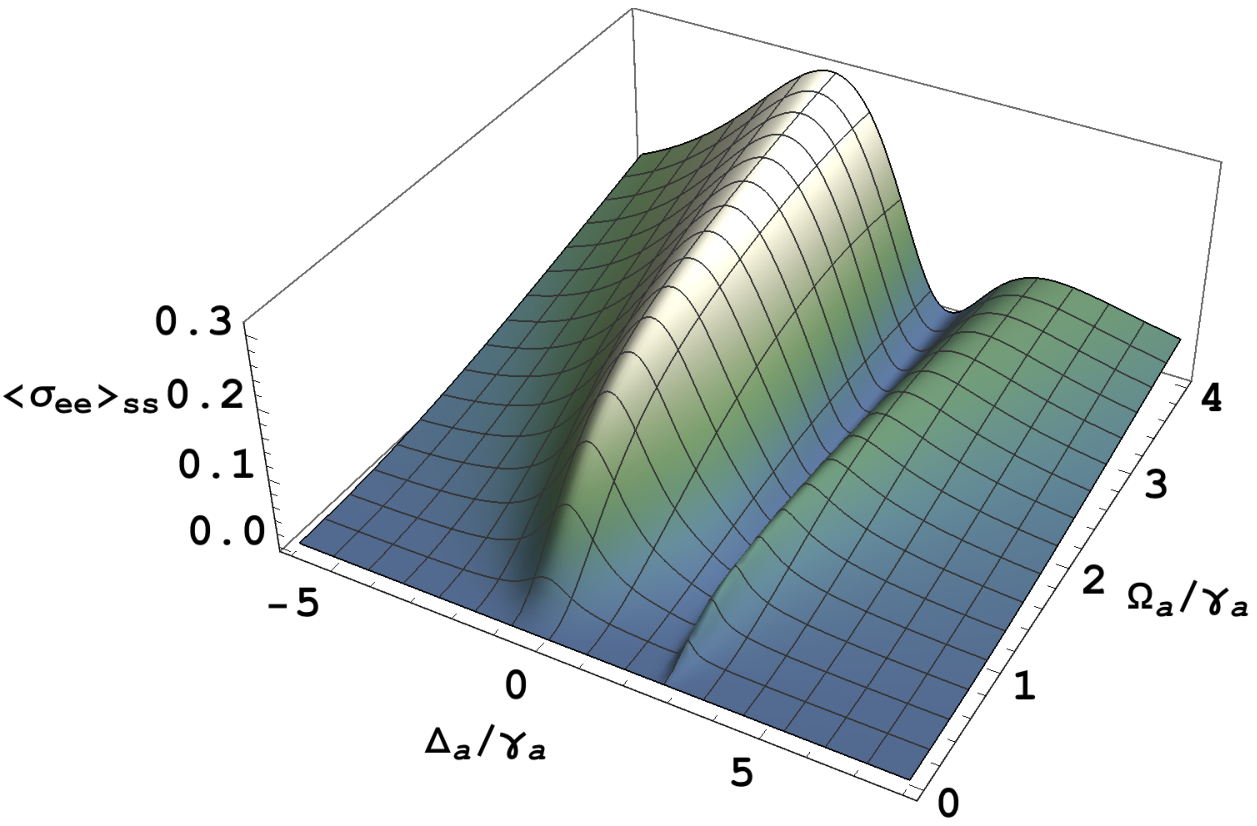} 
\includegraphics[width=8.7cm, height=5.5cm]{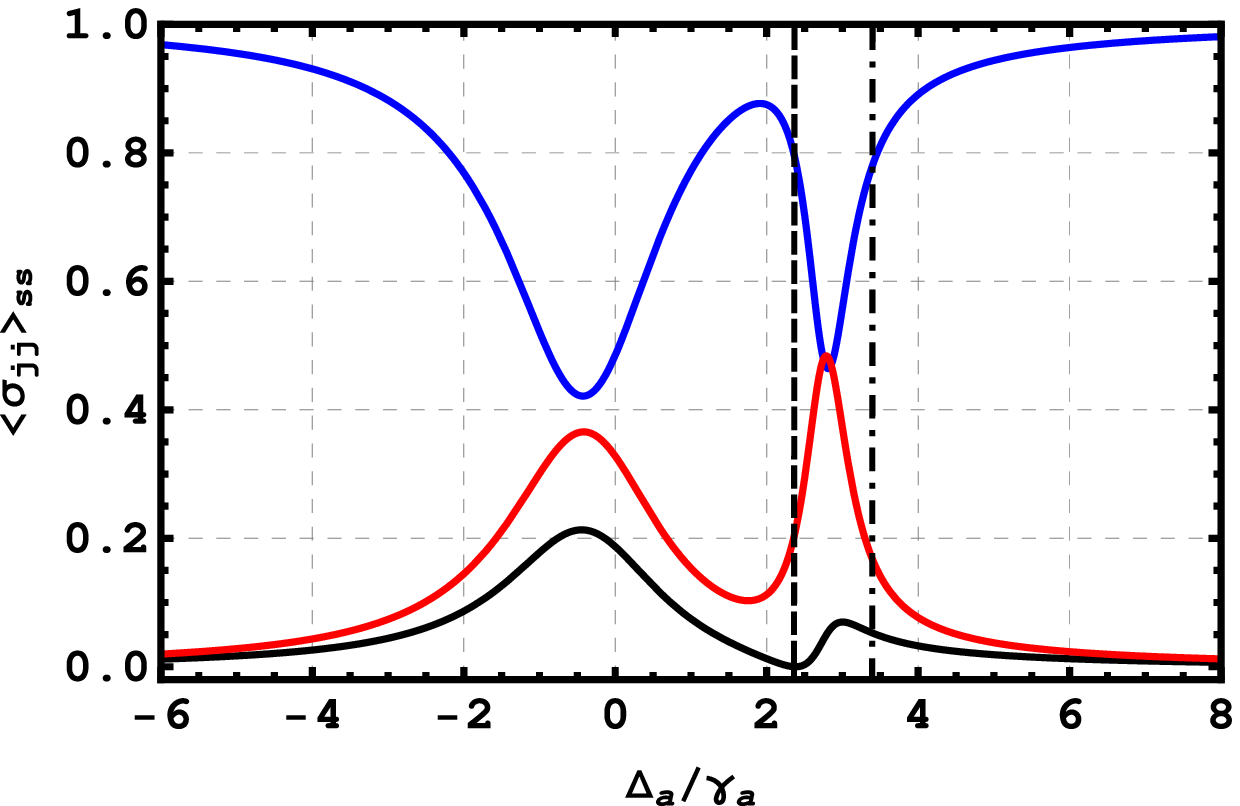} 
\caption{Upper panel: Occupation probability of the excited state, 
$\langle \sigma_{ee} \rangle_{ss}$, versus the scaled detuning 
$\Delta_{a}/\gamma_{a}$ and Rabi frequency $\Omega_{a}/\gamma_{a}$. 
Lower panel: Populations $\langle \sigma_{jj} \rangle_{ss}$, for $j=a ,b$ and 
$e$ (blue, red and black lines, respectively) as functions of 
$\Delta_{a}/\gamma_{a}$; the black curve is the corresponding cross sectional 
profile of the upper figure at $\Omega_{a}/\gamma_{a} \approx 1.12$. The 
remaining parameters are $\Omega_{b}/\gamma_{a}\approx 2.15$ and 
$\Delta_{b}/\gamma_{a}\approx 2.38$.} 
\label{fig:populations}
\end{figure}

Besides, in order for our findings to be possibly put to the test in a given 
realization, we shall consider the decays $\gamma_a =14.7$ MHz and 
$\gamma_b =5.4$ MHz, observed in $^{138} \mathrm{Ba}^+$ ions \cite{GRS+09,SHG+10,DNG+15}. Although a more accurate description of 
Barium resonance fluorescence would entail considering its multilevel structure, 
being composed of eight energy levels, it suffices for our purposes to deal with 
the simplified three-level system as a proxy for specifying the relevant allowed 
dipole transitions that take part in the dynamics. Parenthetically, the isolation of 
a single three-level configuration can be implemented through a proper optical 
pumping arrangement.

\subsection{Role of Coherent Population Trapping}
The $\Lambda$-type three-level atom is an archetypal system that readily fulfills 
the necessary conditions for coherent population trapping (CPT) to take place 
\cite{Arimondo,FlIM05}. In such a scenario, the system is known to evolve towards 
the trapping state $|u \rangle = (\Omega_b |a \rangle - \Omega_a |b \rangle )/ 
\sqrt{ \Omega_a^2 +\Omega_b^2 } $ that turns out to be decoupled from the 
lasers, thereby dropping the long-term excited-state population $\alpha_{ee}$ 
to nearly zero. The manifestation of 
this effect is exemplified in the upper panel of Fig.~\ref{fig:populations}, where the 
steady  state population of the excited state is shown as a function of both the 
detuning and Rabi frequency of the probe laser ($e \to a$ transition); the 
values of the parameters associated with the control field are, henceforth, taken 
to be fixed and the same as those  reported in \cite{SHG+10}, namely, 
$\Omega_{b}/\gamma_{a} \approx 2.15$ and 
$\Delta_{b}/\gamma_{a} \approx 2.38$. In accord with the well-established 
prescription to determine the frequency region around which the atom is 
essentially  transparent to the incoming probe field, the so-called Raman 
resonance condition, the probe detuning must be such that 
$\Delta_{a} \approx \Delta_{b}$ is satisfied; the role of the probe intensity 
$\Omega_{a}$ is that of slightly modifying the width of such a transparent 
frequency window. Its location is also depicted in the lower panel of 
Fig.~\ref{fig:populations} showing the cross sectional profile of the upper figure 
(black line) at $\Omega_{a}/\gamma_{a}\approx 1.12$ where, in turn, we can 
observe the complete depopulation of the excited state at 
$\Delta_{a}/\gamma_{a}\approx 2.38$ (dashed vertical line); the populations of 
the $|a\rangle$ and $|b\rangle$ states are also added as a supplementary view 
of their behavior as functions of the probe detuning.

The foregoing was not the actual condition under which the experiments 
\cite{SHG+10} were performed, but instead the detuning was chosen so as to fit 
the value of the corresponding saturation parameter, 
$\Omega_{j}^{2}/(\gamma_{j}^{2}+\Delta_{j}^{2})$, and taken to be $\sim 0.1$ for 
the $a \to e$ transition. This choice gives  rise to a detuning of about 
$\Delta_{a}/\gamma_{a} \approx 3.4$, the location of which being also indicated 
in the figure (dotted-dashed vertical line); the saturation parameter associated 
with the control field was set to $0.8$. So, for this particular choice of probe and 
control detunings that drive the $a \to e$ transition out of the Raman resonance 
condition, the complete depopulation of the $|e\rangle$ state can be avoided or 
delayed, a working situation that will permit us to study the non-classical properties 
of the scattered light we seek to assess. It is worth commenting that if, instead, the 
$a \to e$ transition were driven more strongly than the $b \to e$ one, such that 
$\Omega_a > \Omega_b$, for general detunings, the population would end up in 
the $|b\rangle$ state, with $\Omega_b \ll \gamma_a$. So then, the strong transition 
would be turned off due to lack of recycling population to $|e\rangle$. \\
\begin{figure}[t!]
\includegraphics[width=8.5cm, height=5cm]{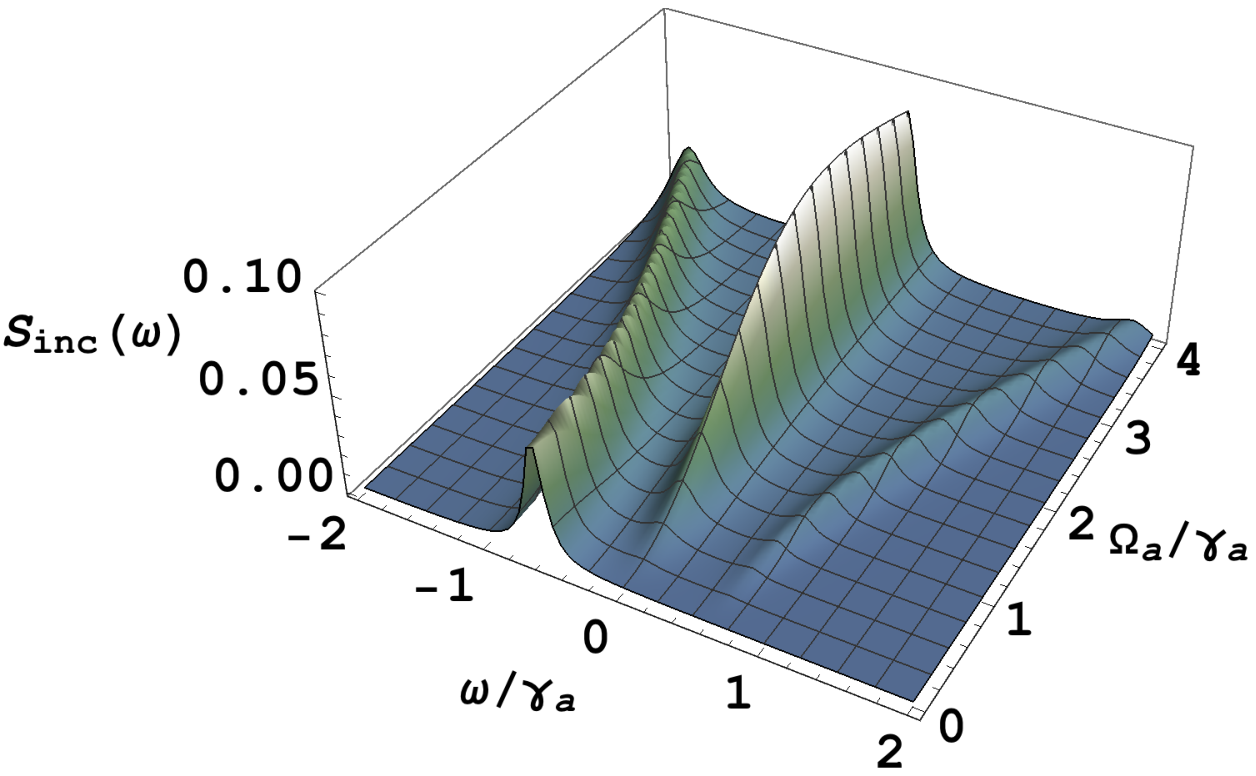} 
\includegraphics[width=8.5cm, height=5cm]{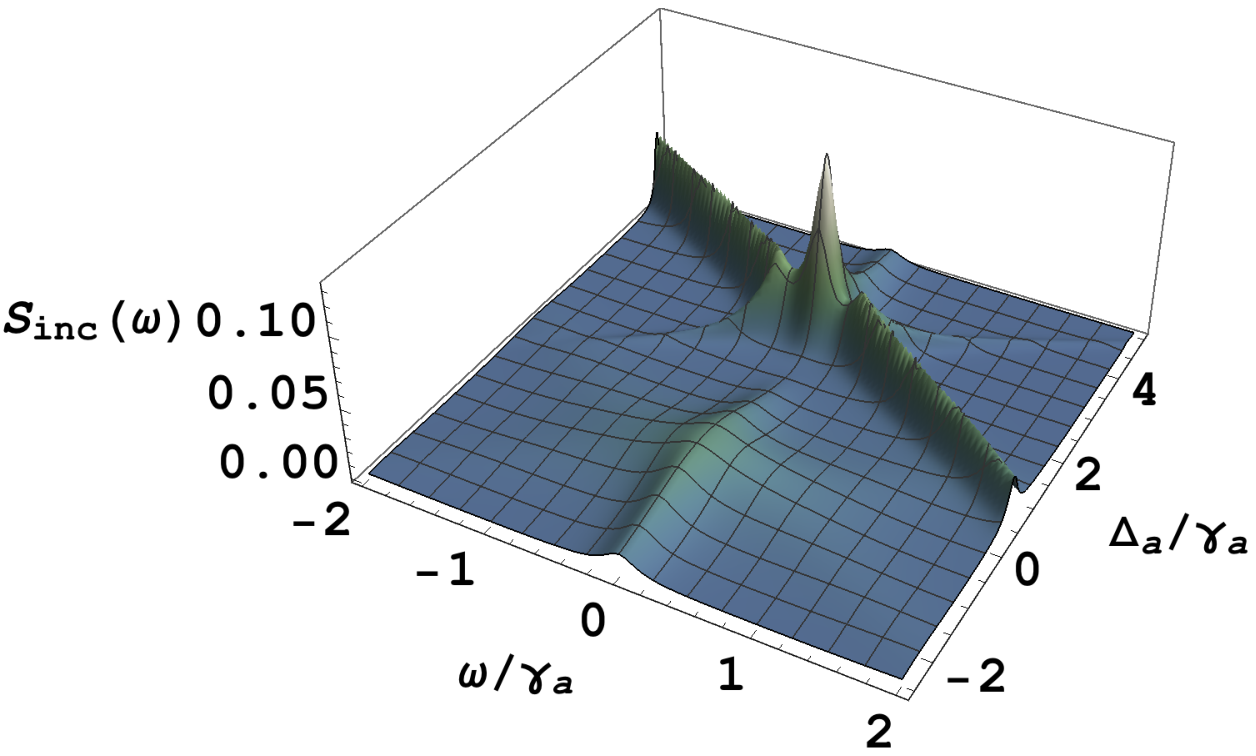}
\caption{Incoherent spectrum, $S_{inc}(\omega)$, of the $e \to a$ transition  
as a function of the scaled probe laser intensity $\Omega_{a}/\gamma_{a}$ 
(upper panel, for $\Delta_{a}/\gamma_{a}\approx 3.4$) and the detuning 
$\Delta_{a}/\gamma_{a}$ (lower panel, for 
$\Omega_{a}/\gamma_{a}\approx 1.12$); the range of spectral distribution is 
also displayed in units of $\gamma_{a}$. The remaining parameters are the 
same as in Fig. \ref{fig:populations}.} 	\label{spectrum_inc}
\end{figure}

So, having established the present configuration of laser intensities and 
frequencies, we find it pertinent, at this stage, to depict the stationary power 
spectrum of the re-emitted light obtained by use of the Wiener-Khintchine 
formula
\begin{equation}
S(\omega) = \frac{1}{\pi \alpha_{ee}} \textrm{Re} \int_{0}^{\infty}d\tau 
 e^{-i\omega \tau} \langle \hat{\sigma}_{ea}(0) \hat{\sigma}_{ae} (\tau) \rangle_{ss},
\end{equation}
i.e., the Fourier transform of the autocorrelation function of the dipole field, 
$\langle \hat{\sigma}_{ea}(0) \hat{\sigma}_{ae} (\tau) \rangle_{ss}$, where $ss$ 
indicates that the process is stationary; the 
prefactor $(\pi \alpha_{ee})^{-1}$ normalizes the integral over all frequencies. 
For convenience, the spectrum is separated into its coherent and incoherent 
parts, namely, $S(\omega)=S_{coh}(\omega)+S_{inc}(\omega)$, as a result of considering the dynamics of the atomic variables to be split into their mean and fluctuations, viz. 
$\hat{\sigma}_{jk} (t) = \alpha_{jk} +\Delta \hat{\sigma}_{jk}(t)$, with 
$\langle \Delta \hat{\sigma}_{jk}(t) \rangle = 0$. In doing so, we get
\begin{equation} 
S_{coh}(\omega) = \frac{|\alpha_{ea}|^{2}}{\pi \alpha_{ee}} 
	\textrm{Re} \int_{0}^{\infty} d \tau e^{-i\omega \tau} 
	= \frac{|\alpha_{ea}|^{2}}{\pi \alpha_{ee}}\delta (\omega),
\end{equation}
where $\alpha_{ea} = \langle \hat{\sigma}_{ea} \rangle_{ss}$, and
\begin{equation}
S_{inc} (\omega) = \frac{1}{\pi \alpha_{ee}} \textrm{Re} \int_{0}^{\infty} 
	d\tau e^{-i\omega \tau} \langle 
	\Delta \hat{\sigma}_{ea}(0) \Delta \hat{\sigma}_{ae}(\tau) \rangle_{ss},
\end{equation}
the former being the coherent constituent of the spectrum owing to elastic 
scattering, and the latter the incoherent part of the spectrum that is brought about 
by atomic fluctuations. The main features of the $\Lambda$-type three-level atom 
spectrum have already been studied from the weak to the strong field limit, both 
theoretically and experimentally \cite{SSA+96}. Figure~\ref{spectrum_inc} shows 
a three dimensional view of the incoherent part of the spectrum associated with 
the $e \to a$ transition, the one  of interest to us, as a function of the probe 
intensity (upper panel) and the detuning (lower panel); details of the steps 
involved in the calculations herein via the matrix analysis are included in 
Appendix \ref{sec:appendix}. The general spectral profile can be understood in 
terms of dressed-state configuration that follows from properly diagonalizing 
the atom-field Hamiltonian \cite{Cohen92}, emphasizing the fact that, above 
saturation, the spectrum displays the appearance of Rabi sidebands as the 
intensity field increases, as one can see in the upper panel of the figure. 
By setting the Rabi frequency at, say, $\Omega_{a}/\gamma_{a}=1.12$, such sidebands become sufficiently conspicuous and the dependency of their profile 
upon the detuning is shown in the lower panel.

\section{Amplitude-Intensity Correlation}
\begin{figure}[t!] 	
\includegraphics[width=8.5cm, height=5.5cm]{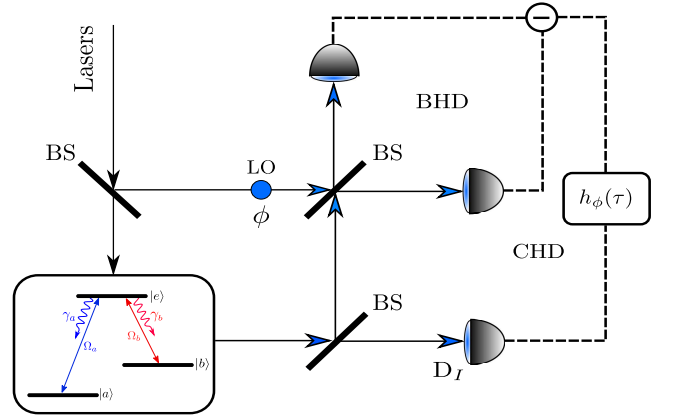}
\caption{Scheme of conditional homodyne detection (CHD). It features a balance 
homodyne detection setup to assess the quadrature of the field on the condition 
of photon detection via detector $\textrm{D}_{I}$. BS and LO stand for beam 
splitter and local oscillator, respectively. }  		\label{fig:chd} 
\end{figure}

In this section we present the theory of conditional homodyne detection (CHD) 
in order to assess and discuss the time-asymmetry, the non-Gaussianity and the 
non-classicality of the light scattered from the atomic system under study. In the 
next section we move to the frequency domain. The CHD setup is sketched in Fig.~\ref{fig:chd}. 
In one arm of the setup a quadrature of the source light, $E_{\phi} \propto 
\hat{\sigma}_{\phi} =(\hat{\sigma}_{ea} e^{-i\phi} +\hat{\sigma}_{ae} e^{i\phi})/2$,  
is analyzed in balanced homodyne detection (BHD), where $\phi$ is the phase 
of the local oscillator (LO). This signal has a delay $\tau$ with respect to the 
measurement of the source's intensity in another arm, proportional to the 
excited-state population, 
$I \propto \langle \hat{\sigma}_{ea} \hat{\sigma}_{ae}\rangle 
= \langle \hat{\sigma}_{ee} \rangle $. Thus, the outcome is an 
amplitude-intensity correlation that reads 
\begin{eqnarray}  	\label{eq:haicdef}
h_{\phi}(\tau) = \frac{\langle: \hat{\sigma}_{ea}(0) \hat{\sigma}_{ae}(0) 
	\hat{\sigma}_{\phi}(\tau): \rangle_{ss} }{ \alpha_{ee} \alpha_{\phi} } \,, 
\end{eqnarray}
where the dots $::$ indicate normal and time operator ordering, and 
$\alpha_{\phi} = (\alpha_{ea} e^{-i\phi} +\alpha_{ae} e^{i\phi})/2$ is the 
stationary value of the quadrature amplitude. 

\subsection{Time-asymmetry and Non-Gaussianity} 	
Resonance fluorescence is a highly non-linear process, preventing its 
description in terms of quasi-probability distributions, i.e., it does not admit 
a Fokker-Planck type of equation. The non-linearity leads to non-Gaussian 
fluctuations, thus giving rise to non-vanishing odd-order moments. Autocorrelation functions such as that for the spectrum, 
$\langle \sigma_{ea}(0) \sigma_{ae} (\tau) \rangle$; squeezing, 
$\langle \Delta \sigma_{\phi}(0) \Delta \sigma_{\phi} (\tau) \rangle$; 
and photon-photon correlation, $\langle \sigma_{ea}(0) \sigma_{ea} (\tau) 
\sigma_{ae} (\tau) \sigma_{ae} (0) \rangle$, are of even-order and, as such, 
time-symmetric \cite{DeCC02}. These functions do not address the 
non-Gaussianity of the field's fluctuations. \\

In amplitude-intensity correlations, Eq.(\ref{eq:haicdef}), on the other hand, 
such a symmetry is not guaranteed: being different observables, the outcome 
will be dependent on the time order of measurements. For instance, the 
quadrature is measured (preselected) for $\tau \geq 0$ and the intensity for 
$\tau \leq 0$ (quadrature is post-selected), a process in which time-asymmetry 
is expected to be revealed. Moreover, this correlation would allow us to explore non-classical features of light beyond squeezing and antibunching. \\

Applying the time and normal operator orderings in Eq.(\ref{eq:haicdef}) we 
arrive at the following expressions for positive and negative time intervals, 
\begin{eqnarray} 
h_{\phi}(\tau \geq 0) 
	&=& \frac{\langle \hat{\sigma}_{ea}(0)  \hat{\sigma}_{\phi}(\tau) 
	\hat{\sigma}_{ae}(0) \rangle_{ss} }{ \alpha_{ee} \alpha_{\phi} } \,, 	\\ 
h_{\phi}(\tau \leq 0) 
	&=& \frac{\mathrm{Re} [e^{-i\phi} \langle \hat{\sigma}_{ea}(0)  
	\hat{\sigma}_{ee}(\tau_-) \rangle_{ss} ]}{ \alpha_{ee} \alpha_{\phi} } \,.  	
\end{eqnarray}
\begin{figure}[t!] 	
\includegraphics[width=8.5cm, height=5cm]{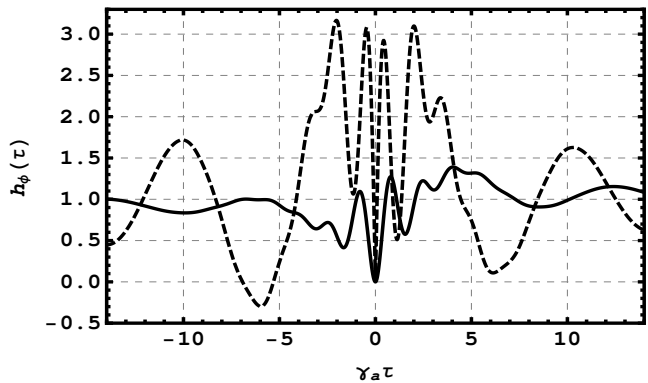} 
\caption{Amplitude-intensity correlations of light from the $e \to a $ transition, 
as a function of the scaled time $\gamma_{a}\tau$, for $\phi=0$ (continuous 
line) and $\phi=\pi/2$ (dotted-dashed line). The parameters are: $\Omega_{a} 
\approx 1.12 \gamma_{a}$, $\Omega_{b} \approx 2.15 \gamma_{a}$
$\Delta_{a} =3.40 \gamma_a$, $\Delta_{b} = 2.38 \gamma_a$.  }  
\label{fig:htau1} 
\end{figure}

The asymmetry in time revealed by CHD, as shown in Fig.~\ref{fig:htau1}, 
is an indicative of non-Gaussian noise. The correlation (\ref{eq:haicdef}) 
contains a product of three dipole operators or, more generally, three field 
amplitude operators. This means that $h_{\phi}(\tau)$ provides access up 
to third order fluctuations; since these are non-Gaussian, this third-order 
correlation does not vanish. To better distinguish the asymmetry and the 
size of these fluctuations, we proceed, as we did with the spectrum, to split 
the dipole dynamics into its mean plus fluctuations, 
$\hat{\sigma}_{jk} = \alpha_{\phi} +\Delta \hat{\sigma}_{jk}$ \cite{hmcb10}, 
\begin{eqnarray} 
h_{\phi}(\tau) &=& 1+ h_{\phi}^{(2)}(\tau) +h_{\phi}^{(3)}(\tau),
\label{eq:h_split}
\end{eqnarray}
where
\begin{equation} 
h_{\phi}^{(2)}(\tau) = \frac{\langle: [\alpha_{ea}  \Delta \hat{\sigma}_{ae}(0) 
+\alpha_{ae} \Delta \hat{\sigma}_{ea}(0)] \Delta \hat{\sigma}_{\phi}(\tau) :\rangle_{ss}}
{ \alpha_{ee} \alpha_{\phi} }, 
\end{equation}
and
\begin{equation}
h_{\phi}^{(3)}(\tau) = \frac{\langle: \Delta \hat{\sigma}_{ea}(0) 
	\Delta \hat{\sigma}_{ae}(0) \Delta \hat{\sigma}_{\phi}(\tau) :\rangle_{ss}} 
	{ \alpha_{ee} \alpha_{\phi} },
\end{equation}
are the components of, respectively, second- and third-order in the dipole 
fluctuations of $h_{\phi}(\tau)$, where $\Delta \hat{\sigma}_{\phi} 
=(\Delta \hat{\sigma}_{ea} e^{-i\phi} +\Delta \hat{\sigma}_{ae} e^{i\phi})/2$ is the 
quadrature fluctuation operator. Fluctuations are said to be Gaussian if 
$h_{\phi}^{(3)}(\tau) \to 0$, which can occur when the transition is weakly driven.  

For positive time intervals between photon and quadrature detections, we get 
\begin{eqnarray} 	
h_{\phi}^{(2)}(\tau \geq 0) &=& \frac{ 2\mathrm{Re} [\alpha_{ae}  
\langle  \Delta \hat{\sigma}_{ea}(0)  \Delta \hat{\sigma}_{\phi}(\tau) \rangle_{st}]}
{ \alpha_{ee} \alpha_{\phi} }, 	\label{eq:htaup2} \\ 
h_{\phi}^{(3)}(\tau \geq 0) &=& \frac{\langle \Delta \hat{\sigma}_{ea}(0) 
\Delta \hat{\sigma}_{\phi}(\tau) \Delta \hat{\sigma}_{ae}(0) \rangle_{st} }
 { \alpha_{ee} \alpha_{\phi} }.  	\label{eq:htaup3}
\end{eqnarray}
\begin{figure}[t!] 	
\includegraphics[width=8.5cm, height=5cm]{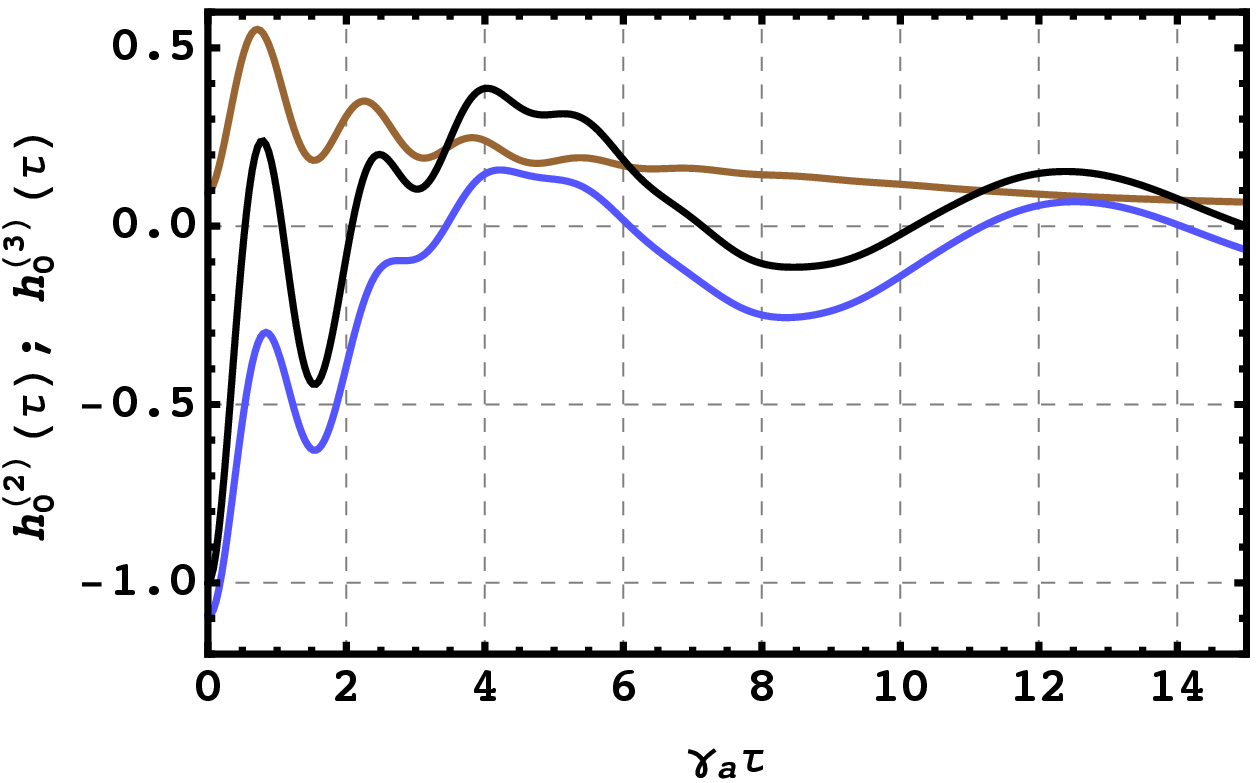} 
\includegraphics[width=8.5cm, height=5cm]{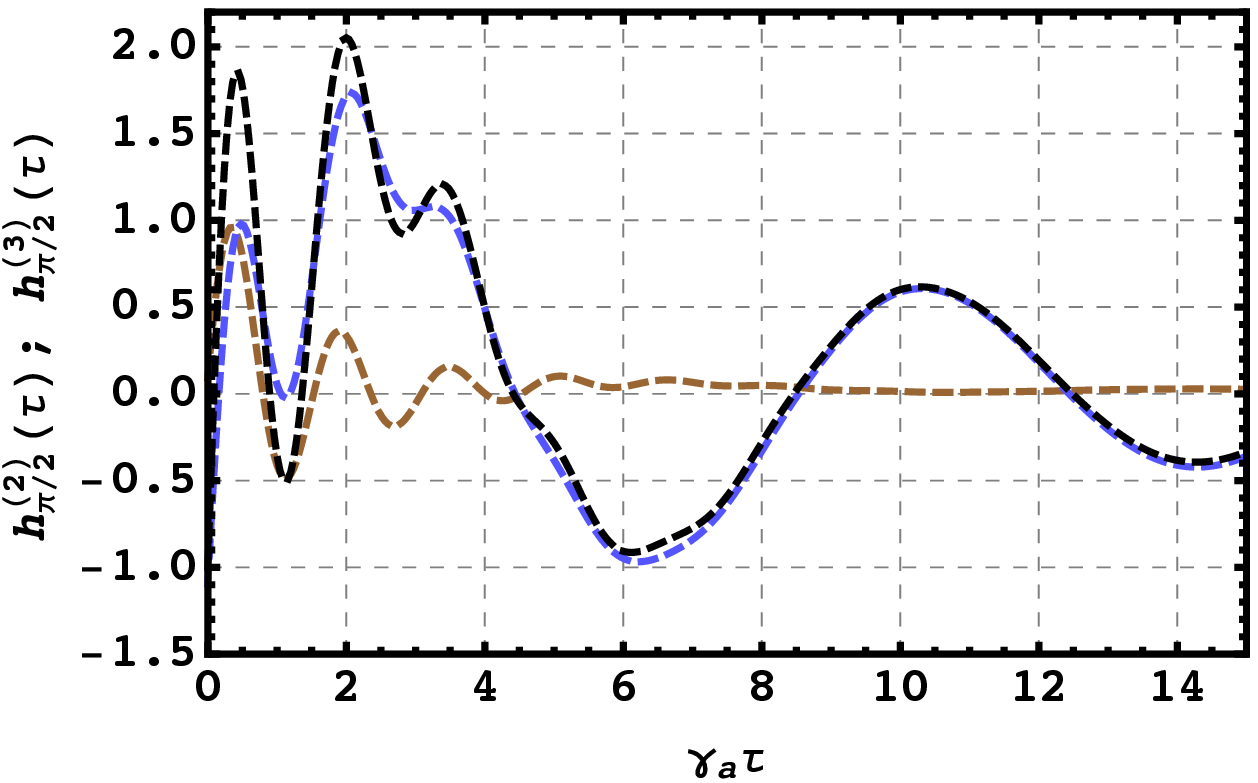} 
\caption{Splitting of the intensity-field correlations shown in Fig.~\ref{fig:htau1} 
into their second- ($h_{\phi}^{(2)}(\tau)$, brown line) and third-order 
($h_{\phi}^{(3)}(\tau)$, blue line) constituents, for $\phi=0$ (upper panel) and 
$\phi=\pi/2$ (lower panel). For comparison, in both cases the black curve 
represents the sum $h_{\phi}^{(2)}(\tau)+h_{\phi}^{(3)}(\tau)$. The parameters 
are the same as those of Fig.~\ref{fig:htau1}. }  		\label{fig:htau2} 
\end{figure}
%
We show in Fig.~\ref{fig:htau2} the foregoing second- and third-order 
correlations. For both quadratures, the third-order constituent (blue lines) 
represents the main contribution, almost that of the total $h_{\phi} (\tau)$ 
(black lines). This is understandable from the fact that we are above the 
saturation threshold \cite{GCRH17}, a regime where non-Gaussian 
fluctuations become significant. In this regime the dipole $\alpha_{ea}$, 
indicative of the coherence induced by the laser, is small; most of the total 
emission is incoherent. This observation can be quantitatively revealed from 
the fact that $h_{\phi}(0) =0$ (just as it occurs for photon correlations in 
resonance fluorescence), which leads to the relation \cite{GCRH17}
\begin{equation} 
h_{\phi}^{(3)}(0) = -\left[ 1+h_{\phi}^{(2)}(0) \right] 
	= \frac{2(|\alpha_{ea}|^2 -\alpha_{ee})}{\alpha_{ee}}.
\end{equation}
For strong fields, $|\alpha_{ea}| \ll \alpha_{ee}$, $h_{\phi}^{(3)}(0)$ 
reaches its extremal value -2, thereby making the dipole factor in 
Eq.(\ref{eq:htaup2}) small compared to the third-order term.  

Even though the splitting itself cannot be directly realizable from the 
experimental viewpoint via the measurement scheme, it provides us with 
valuable theoretical information to be able to discern the actual contribution to 
the system's fluctuations.

For $\tau < 0$, we want to stress the fact that the outcome of CHD correlation 
should be taken with special care: it is to be interpreted as the measurement of 
the intensity \textit{after} the detection of the amplitude. Thus, as previously 
underlined, the asymmetry results from the different fluctuations of the light's 
amplitude and intensity. Time and normal operator ordering leads to
\begin{eqnarray}
h_{\phi} (\tau \leq 0) = 1+  \frac{ \mathrm{Re} [ e^{-i\phi}  
\langle  \Delta \hat{\sigma}_{ea}(0)  \Delta \hat{\sigma}_{ee}(|\tau|) \rangle_{st} ]}
{\alpha_{ee} \alpha_{\phi} } \,,
\end{eqnarray}
i. e., the correlation is only of second order in the dipole fluctuations, albeit 
with  $\Delta \hat{\sigma}_{ee}$ instead of the quadrature amplitude 
$\Delta \hat{\sigma}_{\phi}$ fluctuation operator.

\subsection{Non-classicality} 	
The initial motivation for CHD was to detect squeezing from weak sources, 
such as cavity QED \cite{CCFO00,FOCC00}. Resonance fluorescence is also 
a producer of weakly squeezed light \cite{WaZo81,CoWZ84}. In order to 
produce light in a squeezed state, a non-classical property of light, these 
sources must be weakly driven, so that the third-order fluctuations discussed 
above are small. As we will see later, the remaining second-order signal  is 
related to the spectrum of squeezing. CHD, hence, gives non-classical 
criteria in the time domain as resulting of violation of the classical inequalities 
\cite{CCFO00,FOCC00}
\begin{eqnarray}
0 \le h_{\phi}(\tau)-1 & \le & 1, \\
|h_{\phi}^{(2)}(\tau)| \le |h_{\phi}^{(2)}(0)| & \le & 1, 	
\end{eqnarray}
where the second relation is derived for Gaussian fluctuations. More recently, it was 
found that light in a coherent state obeys \cite{GCRH17}
\begin{eqnarray}
-1 \le h_{\phi}(\tau) & \le & 1;
\end{eqnarray}
light outside these bounds violates Poissonian statistics. According to these 
criteria, we see in Figs.~\ref{fig:htau1} and \ref{fig:htau2} that both the in-phase 
($\phi=0$, continuous line) and out-of-phase ($\phi=\pi/2$, dashed line) 
quadratures of the field display a non-classical character, violating one or more inequalities. The fact that $h_{\phi}(0) =0$ already shows a non-classical 
feature, akin to antibunching in the intensity fluctuations. Also, moderately 
strong fields easily drive $h_{\phi}(\tau)$ out of the classical bounds. 

We see, then, that CHD clearly reveals non-classicality of quadratures in the 
time domain. Let us now proceed to scrutinize the spectral profile of 
amplitude-intensity correlations in the frequency domain. 

\section{Quadrature Spectra} 
\begin{figure}[t!]
\includegraphics[width=8.5cm, height=5.0cm]{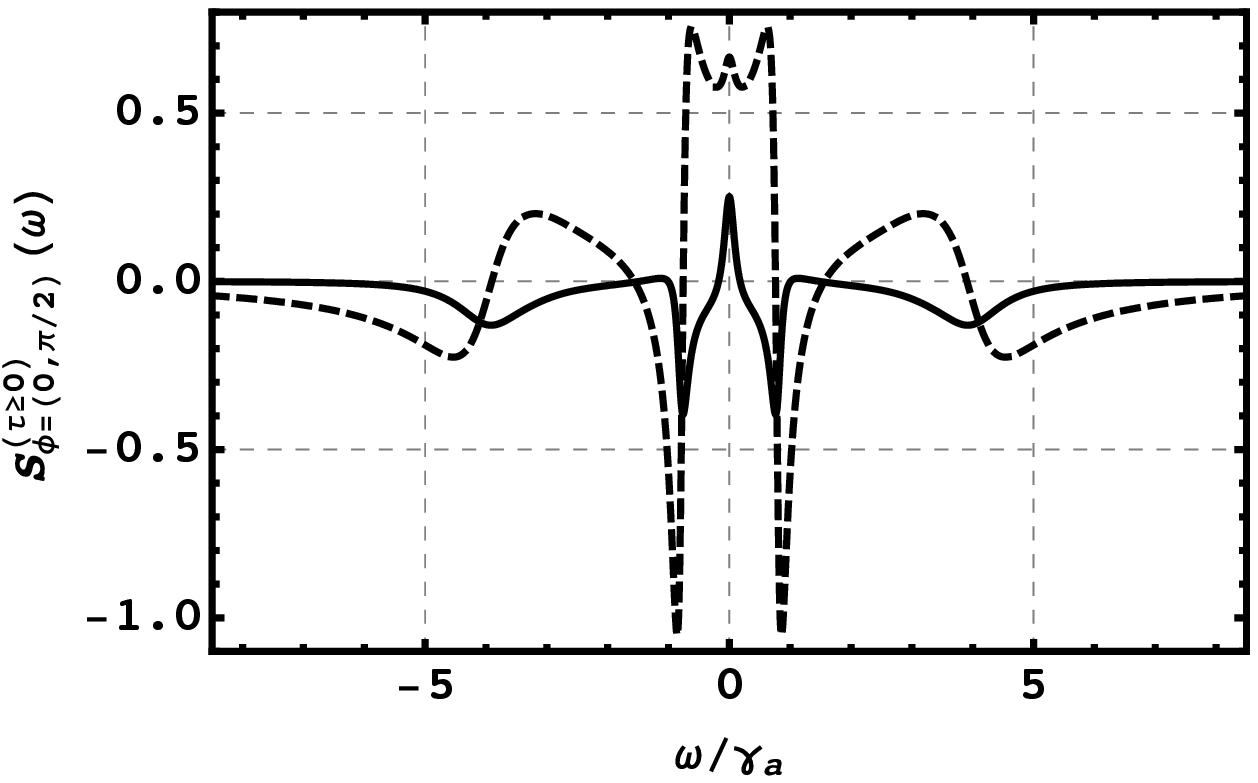} 
\includegraphics[width=8.5cm, height=5.0cm]{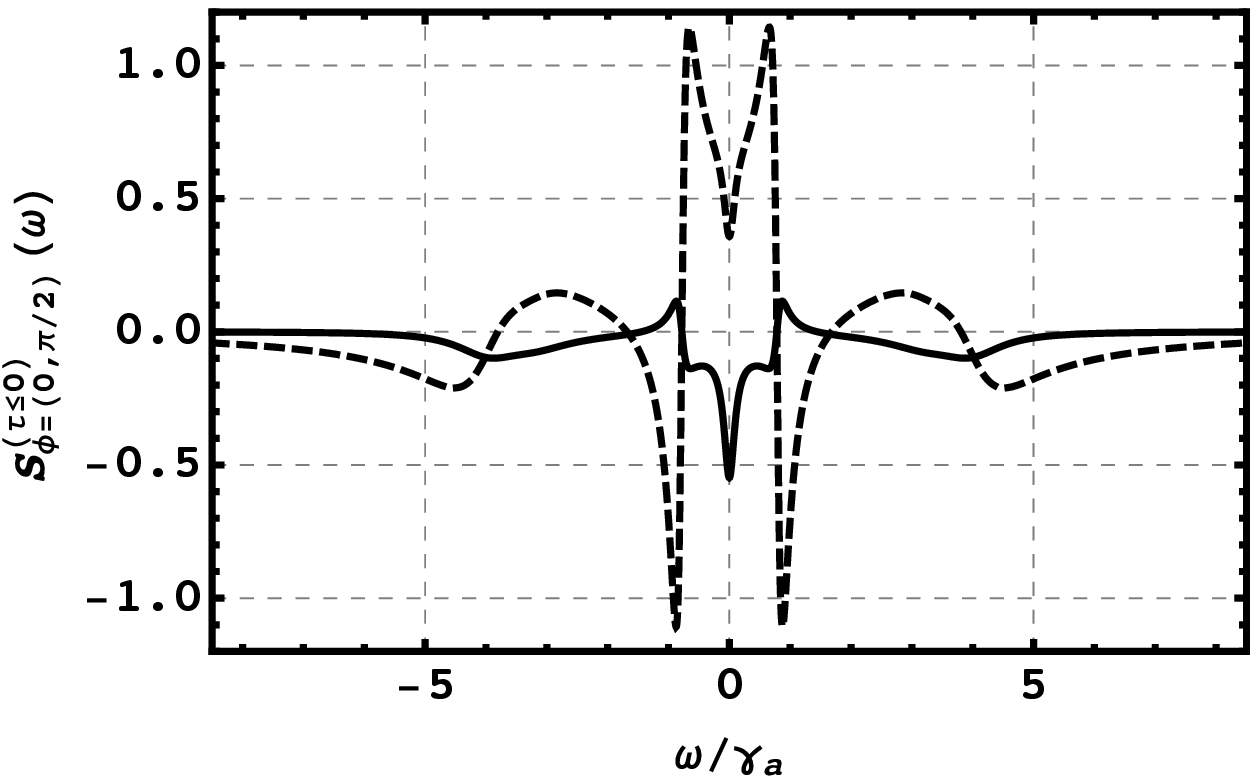} 
\caption{Spectra, Eqs.~(\ref{eq:S-chd}) and (\ref{eq:Sn-chd}), upper and lower 
panels, respectively,  for $\phi=0$ (continuous line) and $\pi/2$ (dashed line). 
The parameters are the same as those of Fig.~\ref{fig:htau1}. }
\label{fig:fullspect}
\end{figure}
\begin{figure}[t!]
\includegraphics[width=8.5cm, height=4.5cm]{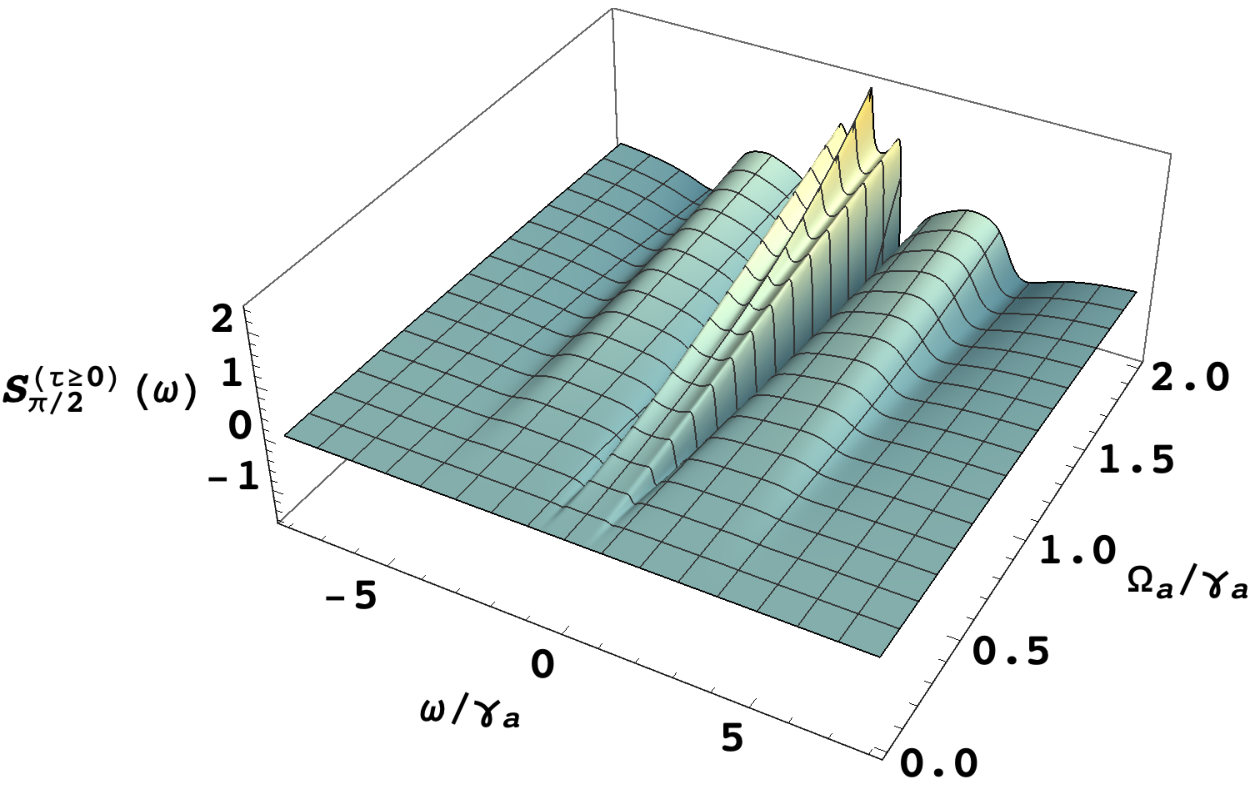}
\includegraphics[width=8.5cm, height=4.5cm]{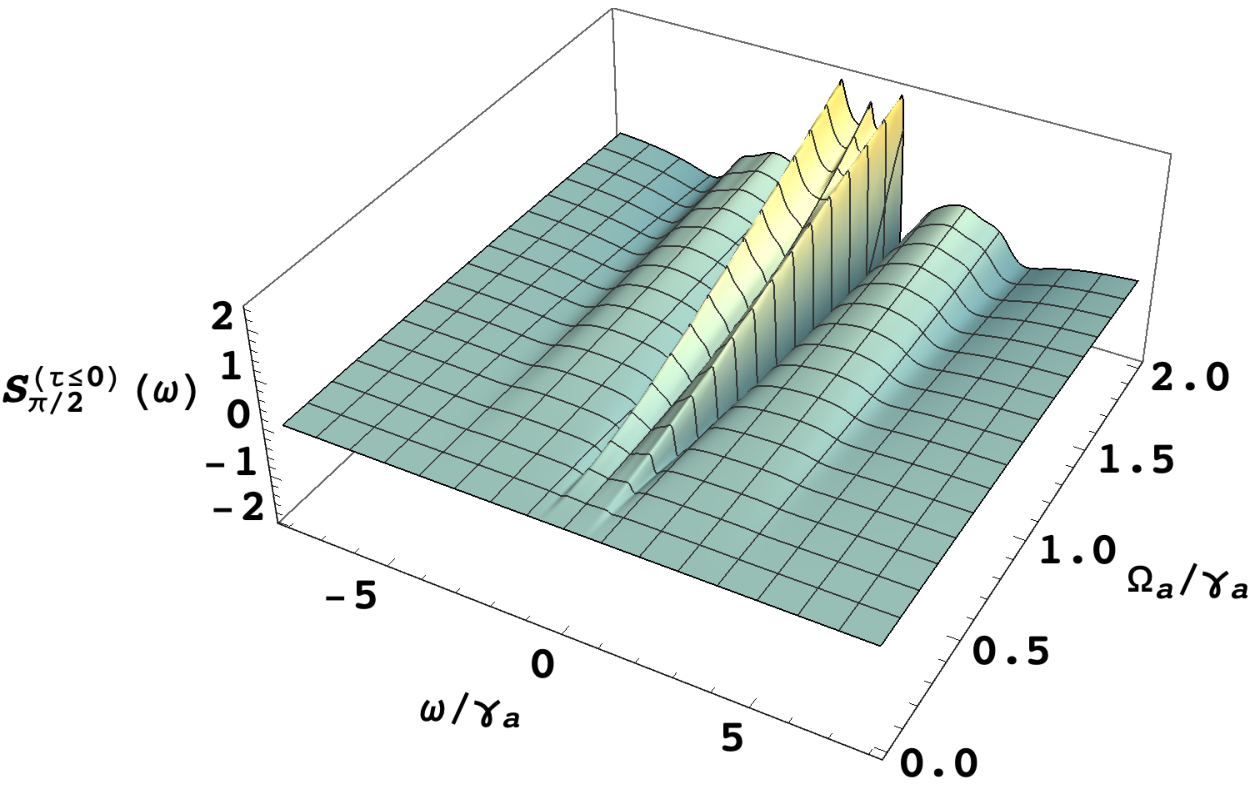} 
\caption{Fourier cosine transform of $h_{\pi/2}^{(N)}(\tau)$, for $\tau \ge 0$ 
(upper panel) and $\tau \le 0$ (lower panel), as a function of the scaled Rabi 
frequency $\Omega_{a}/\gamma_{a}$, for 
$\Omega_{b} \approx 2.15 \gamma_{a}$
$\Delta_{a} =3.4 \gamma_a$ and $\Delta_{b} = 2.38 \gamma_a$. }  		\label{fig:squeezing-2}
\end{figure}

Since in CHD the signal is time-asymmetric, carrying different information for 
positive and negative intervals, the spectra of quadratures measured from 
the amplitude-intensity correlation should be calculated separately 
\cite{GCRH17}: 
\begin{eqnarray}  
S_{\phi}^{(\tau \ge 0)}(\omega) = 4 \gamma_{a}  \alpha_{ee}  \int_{0}^{\infty}  
	d\tau \cos{\omega \tau} \left[ h_{\phi}(\tau \ge 0) -1 \right] 	
	\label{eq:S-chd}   \,, \\ 
S_{\phi}^{(\tau \le 0)} = 4\gamma_{a} \alpha_{ee} \int_{-\infty}^0 d\tau 
	\cos(\omega \tau)[h_{\phi}(\tau \le 0) -1] 	\,, 	\label{eq:Sn-chd}
\end{eqnarray}
for positive and negative time intervals, respectively. The prefactor 
$\gamma_{a}  \alpha_{ee}$ is the photon emission rate in the probe transition. 
In Fig.~\ref{fig:fullspect} we show the spectra calculated from 
Eqs.~(\ref{eq:S-chd}) and (\ref{eq:Sn-chd}) for both quadratures and the same 
parameter values of Fig.~\ref{fig:htau1}. From the CHD viewpoint, negative 
values of the spectrum are signature of non-classical scattered light beyond 
squeezing, which is confirmed for both quadratures, with the $\pi/2$ quadrature 
exhibiting a more pronounced non-classical behavior than the other. 
Fig.~\ref{fig:squeezing-2} also shows the overall spectral profile as a function of 
the probe laser's Rabi frequency for the $\pi/2$ quadrature only, and for positive 
(upper panel) and negative (lower panel) intervals. This more complete 
landscape allows us to verify non-classicality of light revealed by clear-cut 
negative valleys even for excitation above saturation. 

\begin{figure}[t!]
\includegraphics[width=8.5cm, height=5.0cm]{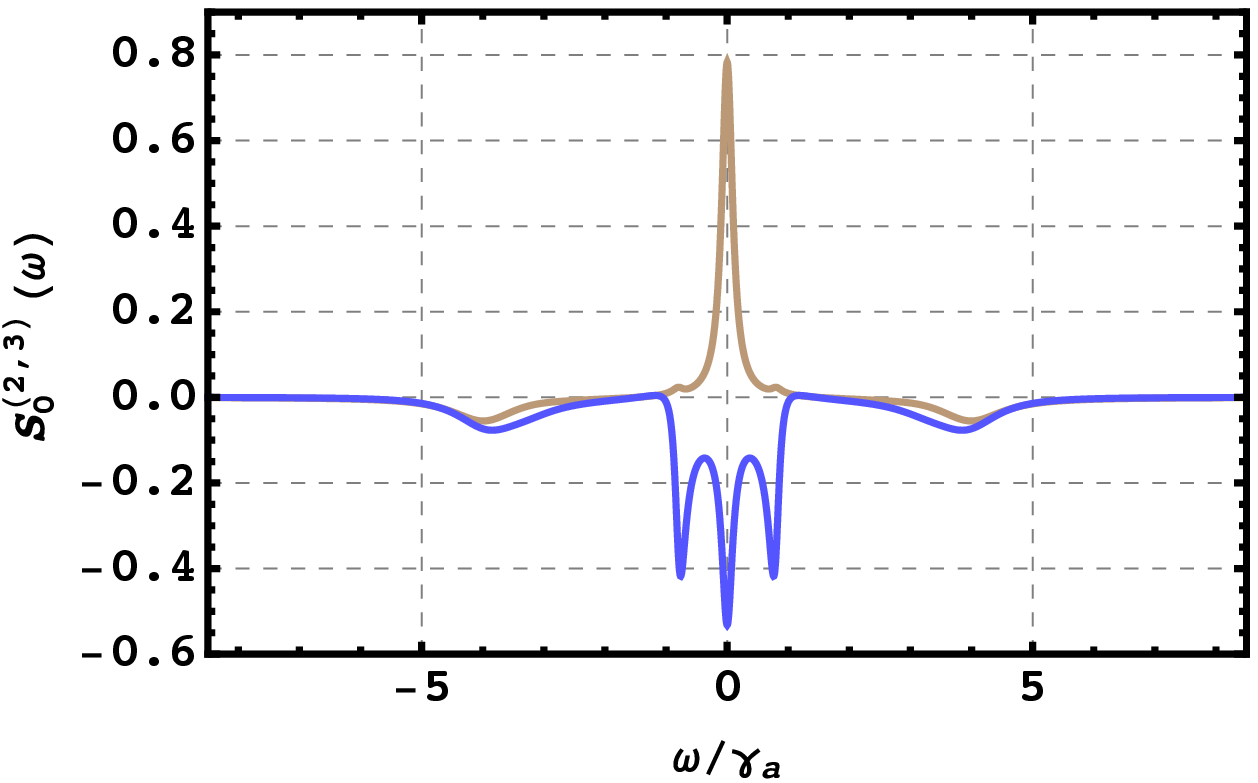} 
\includegraphics[width=8.5cm, height=5.0cm]{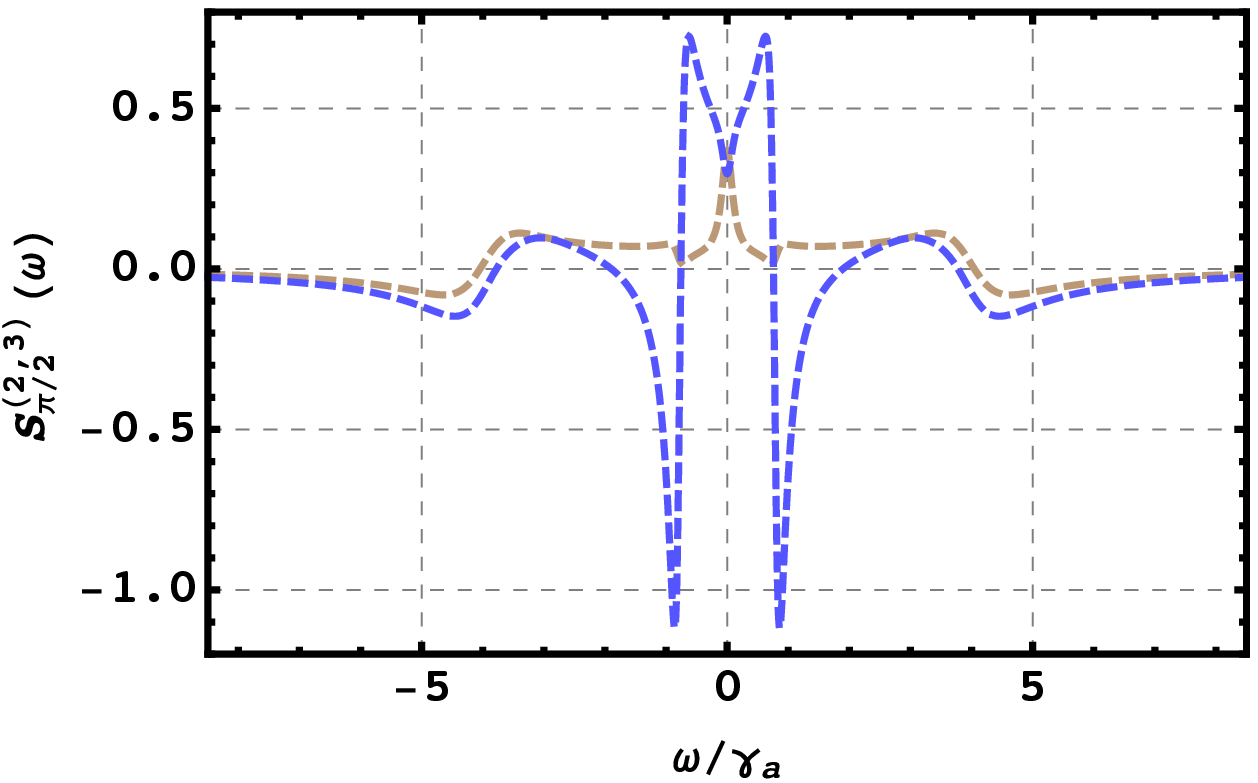} 
\caption{Fourier cosine transform of $h_{\phi}^{(N)}(\tau \ge 0)$, 
Eq.~(\ref{eq:Sk}), for $N=2$ (brown) and $N=3$ (blue). Upper and lower panels 
correspond, respectively, to the cases $\phi=0$ and $\pi/2$. The parameters 
are the same as those of Fig.~\ref{fig:htau1}. } 		\label{fig:splitting}
\end{figure}

\begin{figure}[t!]
\includegraphics[width=8.5cm, height=4.5cm]{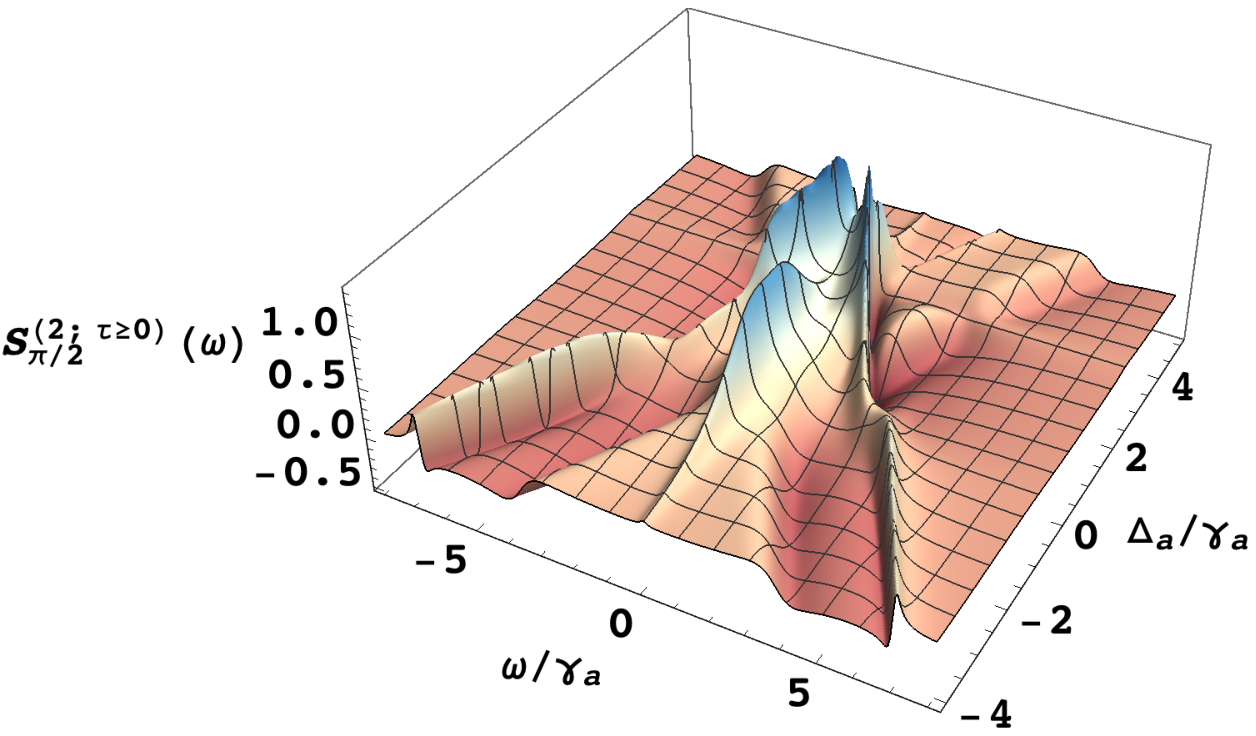}
\includegraphics[width=8.5cm, height=4.5cm]{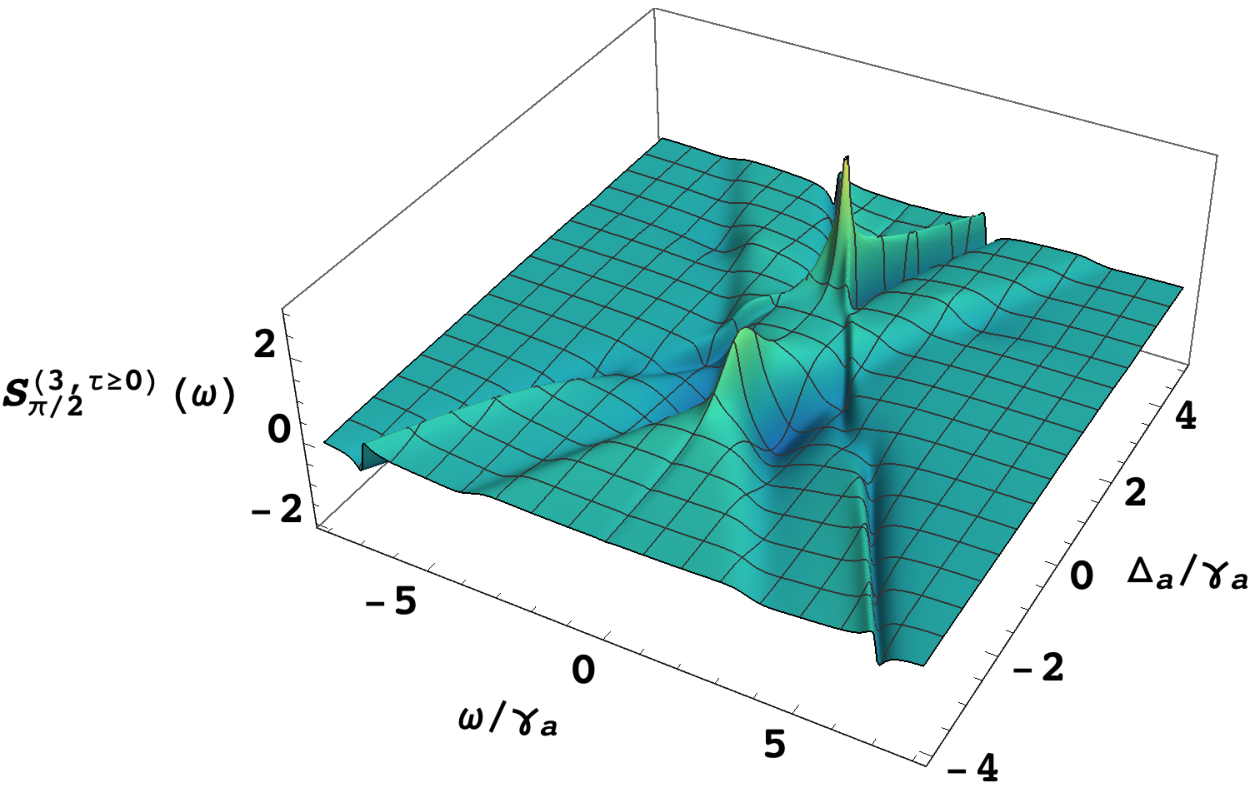} 
\caption{ Fourier transform of $h_{\pi/2}^{(2,3)}(\tau \ge 0)$, as a function of the 
scaled detuning $\Delta_{a}/\gamma_{a}$, for 
$\Omega_{a}/\gamma_{a}\approx 1.12$. The set of parameters is the same 
as in Fig.~\ref{fig:htau1}.  } 		\label{fig:squeezing_del}
\end{figure}

Following the splitting of $h_{\phi}(\tau \ge 0)$, Eq.~(\ref{eq:h_split}), the 
spectra of second- and third-order dipole fluctuations are 
\begin{equation} 	  
S_{\phi}^{(N)}(\omega) = 4\gamma_{a} \alpha_{ee} \int_{0}^{\infty}  d\tau 
\cos{\omega \tau} \ h_{\phi}^{(N)}(\tau),		\label{eq:Sk}
\end{equation}
for $N=2,3$, so that $S_{\phi}^{(\tau \ge 0)}(\omega)=S_{\phi}^{(2)}(\omega) 
+S_{\phi}^{(3)}(\omega)$. These are shown in Fig.~\ref{fig:splitting} for both 
quadratures, corresponding to the CHD signals of Fig.~\ref{fig:htau2}. We 
find that the second-order spectra are mostly positive, while the third-order 
spectrum is negative for $\phi =0$, there are negative 
bands for $\phi =\pi/2$. In Fig.~\ref{fig:squeezing_del} the dependence of $S_{\pi/2}^{(N)}$ 
on the detuning of the probe laser is shown. A quite similar spectral 
landscape (not shown) was found in the second-order correlation for 
$\tau \le 0$, Eq.~(\ref{eq:Sn-chd}). The lineshapes are very complicated, but 
the dispersive features at the sides reveal the non-Gaussianity of the field \cite{CaRG16,GCRH17}. 

The above spectra clearly deviate from the more conventional measure of 
non-classical phase-dependent fluctuations, squeezing, due to the non-linearity 
induced by the strong lasers. Understood operationally as the reduction of 
quantum fluctuations below the shot noise limit, squeezing can be obtained in 
the spectral domain as the Fourier  transform of symmetric photocurrent 
fluctuations in homodyne detection \cite{Carmichael87}. For our source,   
{\setlength\arraycolsep{2pt}
\begin{eqnarray} 	\label{eq:specsqueez} 
S_{\phi}(\omega) &=& 
	8\gamma_{a} \eta \int_{0}^{\infty}  d\tau \cos{\omega \tau} \,	
	 \langle : \Delta \hat{\sigma}_{\phi}(0) \Delta \hat{\sigma}_{\phi}(\tau) : 	
	 \rangle_{ss},  	\nonumber \\
&=& 8\gamma_{a} \eta \int_{0}^{\infty}  d\tau \cos{\omega \tau} 	\nonumber \\
	&& \times \mathrm{Re} \left[ e^{-i\phi}  
	\langle \Delta \hat{\sigma}_{ea}(0) 
	\Delta \hat{\sigma}_{\phi}(\tau) \rangle_{ss}  \right], 
\end{eqnarray}}
where $\eta$ is a combined collection and detection efficiency, and the dots 
$::$ state that the operators must follow time and normal orderings. It was 
shown in \cite{CCFO00,FOCC00} that, in the weak-field limit, when third-order 
fluctuations can be neglected, the second-order spectrum from CHD, from 
Eqs.~(\ref{eq:Sk}) and (\ref{eq:htaup2}), is indeed the spectrum of squeezing, 
but unafected by detector losses, i.e., 
$S_{\phi}^{(2)}(\omega) = S_{\phi}(\omega)/\eta$, owing to the conditional 
character of CHD. 

\begin{figure}[t!]
\includegraphics[width=8.5cm, height=3cm]{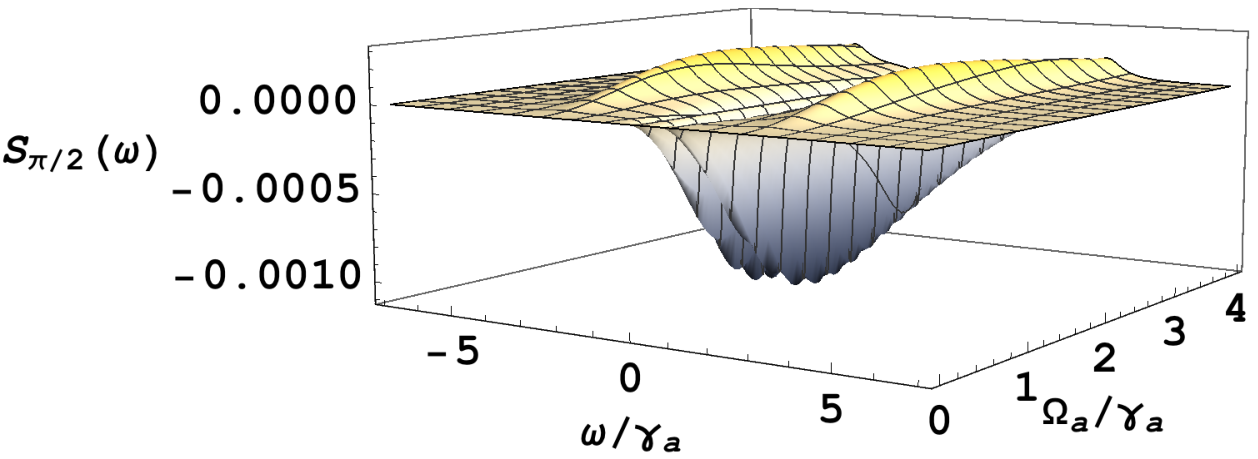}
\includegraphics[width=8.5cm, height=4.6cm]{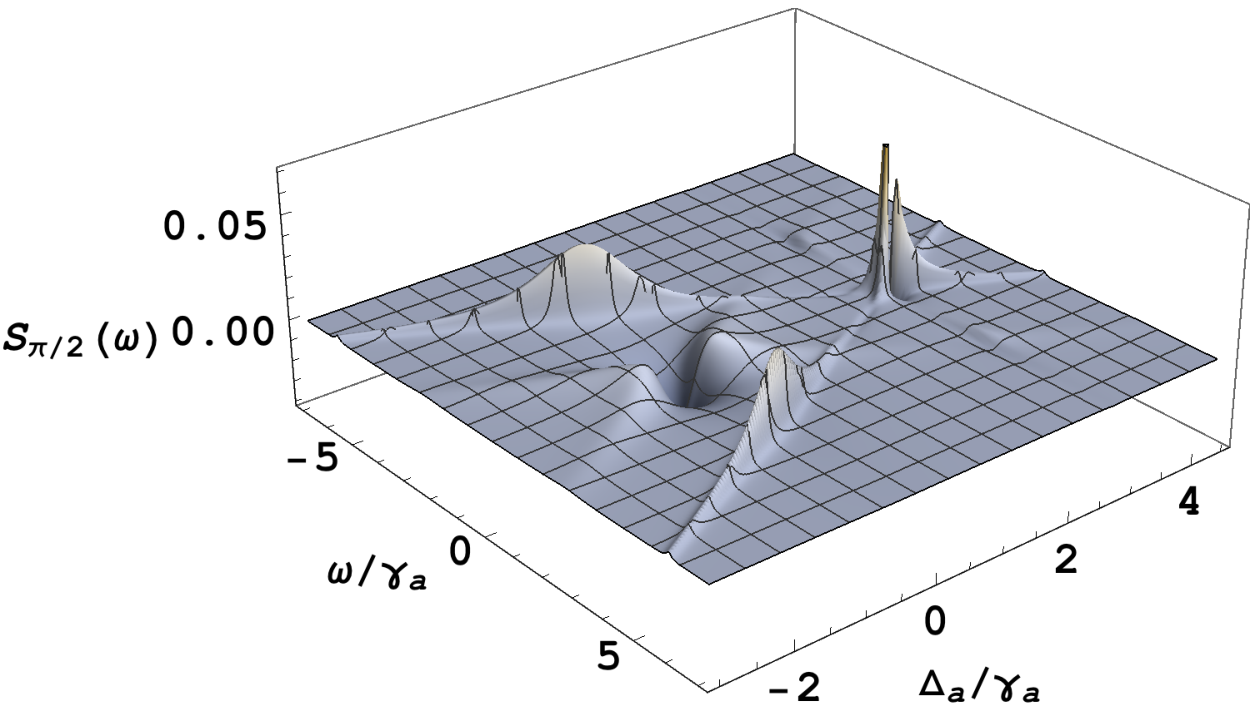} 
\caption{Spectrum of squeezing of the $\phi=\pi/2$ quadrature as a function 
of the probe field intensity (upper panel, for $\Delta_{a}/\gamma_{a}= 2.38$) 
and detuning (lower panel, for $\Omega_{a}/\gamma_{a}=0.1$). For both, 
$\Omega_{b} \approx 2.15 \gamma_{a}$ and $\Delta_{b} = 2.38 \gamma_a$.} 	
\label{fig:specbhd}
\end{figure}

Figure~\ref{fig:specbhd} displays a 3D plot of the spectrum of squeezing, given 
by Eq.~(\ref{eq:specsqueez}) with $\eta =1$, as a function of the probe laser 
intensity $\Omega_{a}/\gamma_{a}$ (upper panel) and detuning 
$\Delta_{a}/\gamma_{a}$ (lower panel), for the $\phi=\pi/2$ quadrature. The 
figure shows up indicatives of squeezing (negative values on the spectral 
content) for a moderate detuning, at $\Delta_{a}/\gamma_{a}=2.38$, around 
which CPT takes place, even for laser intensities above saturation. A slightly 
higher degree of squeezing comes about within certain regions of the 
spectrum by fixing the laser intensity, at $\Omega_{a}/\gamma_{a}=0.1$, say, 
and varying the detuning (see lower panel). 

\begin{figure}[t!]
\includegraphics[width=8.5cm, height=5.5cm]{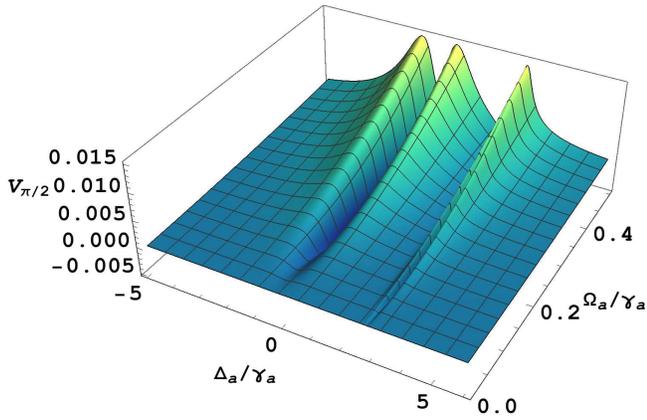}
\caption{Variance for $\phi=\pi/2$, as a function of the scaled Rabi 
frequency $\Omega_{a}/\gamma_{a}$. For both, 
$\Omega_{b} \approx 2.15 \gamma_{a}$ and $\Delta_{b} = 2.38 \gamma_a$.} 
\label{fig:variance}
\end{figure}

An alternative picture of squeezing is the variance 
\begin{eqnarray} 	\label{eq:variance}
V_{\phi} &=& \langle : (\Delta \sigma_{\phi})^2 : \rangle_{st}   
	=\mathrm{Re} \left[ e^{-i\phi}  
	\langle \Delta \sigma_{eg} \Delta \sigma_{\phi} \rangle_{ss}  \right],   
\end{eqnarray}
related to the integrated spectrum as $\int_{-\infty}^{\infty} 
S_{\phi}(\omega) d \omega =4\pi \gamma_a \eta V_{\phi}$. This quantity is 
depicted in Fig.~\ref{fig:variance} for $\phi=\pi/2$ as a function of the scaled 
detuning and laser intensity, revealing a very small degree of squeezing in the quadrature reflected within a restricted region of negative variance. On the 
other hand, the region within which CPT takes hold, around 
$\Delta_{a}/\gamma_{a}\approx 2.38$, is found to reduce fluctuations, 
approximately, to the extent of a coherent state. It was also verified that the 
in-phase quadrature (not shown) did not feature squeezed fluctuations in any 
parameter regime of the aforesaid transition. 

\section{Conclusions}
Using the framework of conditional homodyne detection, we have analyzed the nearby 
effect of coherent population trapping on the phase-dependent quantum 
fluctuations, in both time and frequency domains, of the light fluoresced in the 
probe transition from a coherently driven $\Lambda$-type three-level atom. 
Given the feasibility of implementing the outlined optical system, a single 
$^{138} \mathrm{Ba}^+$ ion \cite{GRS+09,SHG+10}, our findings are expected 
to bolster further experimental investigations to be benchmarked against 
CHD-based theoretical predictions.  

It is worth underlying that the CHD framework proves to be a versatile tool to 
discern the contribution of phase-dependent fluctuations of different orders, 
concluding that the light scattered under the aforesaid conditions is essentially 
non-Gaussian; i.e., the correlation of third-order in the dipole fluctuation operators 
prevails. Non-Gaussianity, notably, manifests in two main ways. On the one hand, 
the amplitude-intensity correlation is, in general, time-asymmetric, indicating that 
amplitude and intensity of the radiated field have different noise properties. 
On the other, the non-linearity imposed by a saturating excitation regime leads to   
fluctuations away from the ideal weak-field squeezing regime. 

The role of CPT in CHD is explored with particular focus on the spectra of 
quadratures. In this regard, as a function of the probe detuning, the spectral 
content confirms once again the prevailing contribution of third-order fluctuations 
to the outcome of the measurements, for both quadratures. This fact is also 
reinforced by examining the variance (the integrated spectrum) of fluorescence. 

\section{Acknowledgments}
The authors thank Dr. Ir\'an Ramos-Prieto for useful coversations and help 
with the figures.

\appendix 

\section{Correlations and Spectra} \label{sec:appendix}
Here, we succinctly describe the evaluation of the expectation values of two-time 
correlations and spectra used throughout this work. 

From the equations of motion of the atomic operators, Eqs.~(\ref{eq:pop1}) to (\ref{eq:coh4}), which can be put into the concise form 
$\dot{\mathbf{s}}(t) = \mathbf{M} \mathbf{s} (t)$, with 
$\mathbf{s} = \left\{ \sigma_{ee}, \sigma_{ae}, \sigma_{be}, \sigma_{ea}, 
\sigma_{aa}, \sigma_{ba}, \sigma_{eb}, \sigma_{ab}, \sigma_{bb} \right\}^{T}$ 
and $\mathbf{M}$ the parameter matrix (to be specified), together with the use 
of the quantum regression formula \cite{Carm99}, we seek the general solution 
to the equation  
\begin{equation}
\partial_{\tau} \mathbf{g}(\tau) = \mathbf{M} \mathbf{g}(\tau),
\end{equation}
where $\mathbf{g}(\tau) = \langle \Delta \sigma_{ea}(0) \Delta \mathbf{s}(\tau) 
\Delta A_-(0) \rangle_{ss}$ is the corresponding vector of correlation functions. 
For the second-order correlations, $\Delta A_- =\mathbf{1}$; for the 
third-order ones, $\Delta A_- =\Delta \sigma_{ae}$. Its solution can be written in 
the form $\mathbf{g}(\tau) = e^{ \mathbf{M} \tau } \mathbf{g}(0)$, where the 
initial condition $\mathbf{g}(0)=\mathbf{g}_{ss}$, given in terms of the steady 
state solution of populations and coherences, is solved numerically. 

The incoherent spectrum requires, for instance, handling the time dependence 
of the correlation $\{ \mathbf{g} (\tau) \}_{m} = \langle \Delta \sigma_{ea}(0) 
\Delta \sigma_{ae}(\tau) \rangle_{ss}$, where the subindex $m$-th denotes the 
element of the vector to be taken. The present matrix analysis saves the work 
of solving the correlation explicitly followed by time integration, namely, for 
$\Delta A_{-}=1$,
\begin{eqnarray} 	
S_{inc}(\omega) 
&=& \frac{1}{ \pi \alpha_{ee}  } \mathrm{Re}   \left \{    \int_0^{\infty} 
	d\tau e^{-(i \omega \mathbf{1} -\mathbf{M}) \tau} 	 
	\langle \Delta \sigma_{ea} \Delta \mathbf{s} \rangle_{ss} \right \}_m   
	\nonumber \\ 
&=& \frac{1}{ \pi \alpha_{ee}  } \mathrm{Re} \left \{ 
	-(i \omega \mathbf{1} -\mathbf{M})^{-1} 
	e^{-(i \omega \mathbf{1} -\mathbf{M}) \tau} \left. \right|_0^{\infty}  
	 \right. 	\nonumber \\
&&	\left. \times 
	\langle \Delta \sigma_{ea} \Delta \mathbf{s} \rangle_{ss} \right \}_m 	
	\nonumber \\ 
&=& \frac{1}{ \pi \alpha_{ee}  } \mathrm{Re} \left \{ 
	(i \omega \mathbf{1} -\mathbf{M})^{-1} 
	\langle \Delta \sigma_{ea} \Delta \mathbf{s} \rangle_{ss} \right \}_m , 	
	\nonumber 
\end{eqnarray}
where $\mathbf{1}$ is the $n \times n$ identity matrix. The spectra 
corresponding to the CHD correlations are calculated in the same manner, 
thus giving us the sought results   
\begin{widetext}     
{\setlength\arraycolsep{2pt}
\begin{eqnarray}
S_{\phi}^{(2)}(\omega) &=&  \frac{2\gamma_{a}}{\alpha_{\phi}} 
\textrm{Re} \left \{ \alpha_{ae}e^{-i\phi}\left[ (i\omega \mathbf{1}-\mathbf{M})^{-1} 
	-(i\omega \mathbf{1}+\mathbf{M})^{-1})\mathbf{g}(0) \right]_{m}  \right \} 
	\nonumber \\
&& + \frac{2\gamma_{a}}{\alpha_{\phi}} \textrm{Re} \left \{ \alpha_{ae} e^{i\phi} 
	\left[ (i\omega \mathbf{1}-\mathbf{M})^{-1} 
	-(i\omega \mathbf{1}+\mathbf{M})^{-1}) \mathbf{g}(0) \right]_{n}  \right \}, 	
	\nonumber \\
S_{\phi}^{(3)}(\omega) &=& \frac{2\gamma_{a}}{\alpha_{\phi}} 
	\textrm{Re} \{ e^{-i\phi} \left[ (i\omega \mathbf{1}-\mathbf{M})^{-1} 
	-(i\omega \mathbf{1}+\mathbf{M})^{-1}) \mathbf{g}(0) \right]_{p} \}, \nonumber 
\end{eqnarray}}
for the second- and third-order fluctuations, respectively, and 
\begin{equation*}
S_{j,\phi}^{(\tau \le 0)}(\omega) =  \frac{2\gamma_{j}}{\alpha_{\phi}} 
	\textrm{Re} \{ e^{-i \phi} [((i\omega \mathbf{1}-\mathbf{M})^{-1} 
	-(i\omega \mathbf{1}+\mathbf{M})^{-1})\mathbf{g}(0)]_{q} \}
\end{equation*}
for fluctuations associated with negative time intervals. The elements of the 
vectors, denoted by subindexes $m,n,p$ and $q$, have to be chosen 
appropriately to match the corresponding correlation it seeks to assess.  

For the sake of completeness, the initial conditions of the correlations ($\tau=0$) 
are encapsulated by using the fluctuation operator approach as
\begin{eqnarray*} 
\langle \Delta \sigma_{ij} \Delta \sigma_{kl} \rangle_{ss} 
 	&=& \alpha_{il} \delta_{jk} -\alpha_{ij} \alpha_{kl}, 	 \\ 	
 \langle \Delta \sigma_{ig} \Delta \sigma_{jk} \Delta \sigma_{gi} \rangle_{ss} 
&=& 2 |\alpha_{ig}|^2 \alpha_{jk}  
	+\alpha_{ii} (\delta_{gj} \delta_{kg} -\alpha_{jk})  
	 -\alpha_{ik} \alpha_{gi} \delta_{gj} 
	-\alpha_{ig} \alpha_{ji} \delta_{kg} \,,
\end{eqnarray*}
for the second- and third-order fluctuations, respectively. More explicitly, for the 
$e \to a$ transition, they become  
\begin{eqnarray*}
\langle \Delta \sigma_{ea} \Delta \mathbf{s} \rangle_{ss}  
&=& \left\{-\alpha_{ea} \alpha_{ee}, \alpha_{ee} -\alpha_{ea} \alpha_{ae}, 
	-\alpha_{ea} \alpha_{be}, 		
	 -\alpha_{ea}^2, \alpha_{ea}(1- \alpha_{aa}),  -\alpha_{ea} \alpha_{ba} 	
	 -\alpha_{ea} \alpha_{eb},  \alpha_{eb}-\alpha_{ea}\alpha_{ab}, 
	-\alpha_{ea} \alpha_{bb}  \right\}^{T}, 
\end{eqnarray*}
and
%
\begin{eqnarray} 
\left( \begin{array}{c} 
\langle \Delta \sigma_{ea} \Delta \sigma_{ee} \Delta \sigma_{ae} \rangle_{ss} \\ 
\langle \Delta \sigma_{ea} \Delta \sigma_{ae} \Delta \sigma_{ae} \rangle_{ss} \\ 
\langle \Delta \sigma_{ea} \Delta \sigma_{be} \Delta \sigma_{ae} \rangle_{ss} \\ 
\langle \Delta \sigma_{ea} \Delta \sigma_{ea} \Delta \sigma_{ae} \rangle_{ss} \\ 
\langle \Delta \sigma_{ea} \Delta \sigma_{aa} \Delta \sigma_{ae} \rangle_{ss} \\
\langle \Delta \sigma_{ea} \Delta \sigma_{ba} \Delta \sigma_{ae} \rangle_{ss} \\
\langle \Delta \sigma_{ea} \Delta \sigma_{eb} \Delta \sigma_{ae} \rangle_{ss} \\ 
\langle \Delta \sigma_{ea} \Delta \sigma_{ab} \Delta \sigma_{ae} \rangle_{ss} \\ 
\langle \Delta \sigma_{ea} \Delta \sigma_{bb} \Delta \sigma_{ae} \rangle_{ss} 
	 \end{array} \right)
&=& \left( \begin{array}{c}  
    \alpha_{ee} (2 |\alpha_{ea}|^2 -\alpha_{ee} ) 		\\  
  -2 \alpha_{ae} (\alpha_{ee}-|\alpha_{ea}|^{2}) 		 \\ 
\alpha_{be} (2 |\alpha_{ea}|^2 -\alpha_{ee} )   		\\  
  2\alpha_{ea} ( |\alpha_{ea}|^2 -\alpha_{ee} ) 		  \\   
  ( 2|\alpha_{ea}|^2 -\alpha_{ee} ) (\alpha_{aa} -1)   	\\
   \alpha_{ba} ( 2|\alpha_{ea}|^2 -\alpha_{ee} ) -\alpha_{ea} \alpha_{be}  \\
  \alpha_{eb} ( 2|\alpha_{ea}|^2 -\alpha_{ee} ) 		\\
   \alpha_{ab} ( 2|\alpha_{ea}|^2 -\alpha_{ee} ) 	-\alpha_{eb} \alpha_{ae}	 \\ 
  \alpha_{bb} (2 |\alpha_{ea}|^2 -\alpha_{ee} ) 		 
  	    \end{array}  \right). \nonumber
\end{eqnarray}
\end{widetext}



\begin{references} 


%
\bibitem{Arimondo} E. Arimondo, Progress in Optics, \textbf{35}, 257 (1996). 
%
\bibitem{FlIM05} M.~Fleischhauer, A.~Imamoglu and J.~P.~Marangos, 
Rev. Mod. Phys. \textbf{77} 633 (2005).
%
\bibitem{SHG+10} L.~Slodi\v{c}ka, G.~H\'etet, S.~Gerber, M.~Hennrich, and 
R.~Blatt, Phys. Rev. A \textbf{105} 153604 (2010). 
%
\bibitem{HPL+09} J.~Hwang, M.~Pototschig, R.~Lettow, G.~Zumofen, A.~Renn, 
S.~G\"otzinger, and V.~Sandoghdar, Nature (London) \textbf{460}, 76 (2009).
%
\bibitem{WaZo81} D.~F. Walls and P. Zoller, Phys. Rev. Lett. \textbf{47} 
	709 (1981).  
%
\bibitem{CoWZ84} M.~J.~Collett, D.~F.~Walls, P.~Zoller, 
 Optics Commun. \textbf{52}, 145-149 (1984). 
%
\bibitem{SHJ+15} C.~H.~H.~Schulte, J.~Hansom, A.~E.~Jones, 
C.~Matthiesen, C.~Le~Gall, and M.~Atat\"ure, Nature \textbf{525}, 222 (2015). 
%
\bibitem{Vogel91} W. Vogel, Phys. Rev. Lett. \textbf{67}, 2450-2452 (1991).
%
\bibitem{Vogel95} W. Vogel, Phys. Rev. A \textbf{51}, 4160 (1995).
%
\bibitem{KVM+17} B.~K\"uhn, W.~Vogel, M.~Mraz, S.~K\"ohnke, and 
B.~Hage, Phys. Rev. Lett. \textbf{118}, 153601 (2017).  
%
\bibitem{GRS+09} S.~Gerber, D.~Rotter, L.~Slodi\v{c}ka, J.~Eschner, 
H.~J.~Carmichael, and R.~Blatt, Phys. Rev. Lett. 
\textbf{102}, 183601 (2009).
%
\bibitem{ScVo05} E.~V.~Shchukin and W.~Vogel, Phys. Rev. A 
\textbf{72}, 043808 (2005). 
%
\bibitem{ScVo06} E.~V.~Shchukin and W.~Vogel, Phys. Rev. Lett. 
\textbf{96}, 200403 (2006).
\bibitem{CCFO00} H.~J. Carmichael, H.~M. Castro-Beltran, G.~T. Foster, 
and L.~A. Orozco, Phys. Rev. Lett. \textbf{85}, 1855 (2000).
%
\bibitem{FOCC00} G.~T. Foster, L.~A. Orozco, H.~M. Castro-Beltran, 
and H.~J. Carmichael, Phys. Rev. Lett. \textbf{85}, 3149 (2000).
%
\bibitem{CFO+04} For a review on CHD see H.~J.~Carmichael, 
G.~T.~Foster, L.~A.~Orozco, J.~E.~Reiner, and P.~R.~Rice, 
 in Progress in Optics 46, E. Wolf, ed. (Elsevier, 2004).
%
\bibitem{hmcb10} H.~M. Castro-Beltran, Opt. Commun. \textbf{283}, 
4680 (2010). 
%
\bibitem{CaGH15} H.~M. Castro-Beltran, L. Gutierrez, and L. Horvath, 
Appl. Math. Inf. Sci. \textbf{9}, 2849 (2015).
%
%
\bibitem{DeCC02} A. Denisov, H.~M. Castro-Beltran, and H.~J. Carmichael, 
Phys. Rev. Lett. \textbf{88}, 243601 (2002). 
%
\bibitem{CaRG16} H.~M. Castro-Beltran, R. Roman-Ancheyta, and 
L. Gutierrez, Phys. Rev. A, \textbf{93}, 033801 (2016).
%
\bibitem{MaCa08} E.~R. Marquina-Cruz and H.~M. Castro-Beltran, 
Laser Phys. \textbf{18}, 157 (2008). 
%
\bibitem{GCRH17} L. Gutierrez, H.~M. Castro-Beltran, R. Roman-Ancheyta, 
and L. Horvath, J. Opt. Soc. Am. B \textbf{34}, 2301 (2017).  
%
\bibitem{XGJM15} Q.~Xu, E.~Greplova, B.~Julsgaard, and K.~M\o lmer,  
Phys. Scripta \textbf{90}, 128004 (2015). 
%
\bibitem{XuMo15} Q.~Xu and K.~M\o lmer, Phys. Rev. A \textbf{92}, 
033830 (2015).
%
\bibitem{GaJM13} S.~Gammelmark, B.~Julsgaard, and K.~M\o lmer, 
 Phys. Rev. Lett. \textbf{111}, 160401 (2013).
%
\bibitem{WaFO16} F.~Wang, X.~Feng, and C.~H.~Oh, 
 Laser Phys. Lett. \textbf{13}, 105201 (2016).
%
\bibitem{Santos19} O. de los Santos-S\'anchez. Front. Phys. \textbf{14}, 
61601 (2019). 
%
\bibitem{Zhao+20} T.~Zhao, Z.-A.~Peng, G.-Q.~Yang, G.-M.~Huang, and 
G.-X.~Li, Optics Express \textbf{28}, 379 (2020). 
%
\bibitem{DNG+15} E.~A. M. Nu\~nez-Portela, A.~T. Grier, K. Jungmann, 
A. Mohanty, N. Valappol, and L. Willmann, Phys. Rev. A, \textbf{91}, 
060501(R) (2015). 
%
\bibitem{SSA+96} Y. Stalgies, I. Siemers, B. Appasamy, T. Altevogt, and 
P. E. Toscheck, Europhys. Lett. \textbf{35}, 259 (1996). 
%
\bibitem{Cohen92} C. Cohen-Tannoudji, J. Dupont-Roc, and G. Grynberg, 
\textit{Atom-Photon Interactions: Basic Processes and Applications} 
(Wiley, New York, 1992).
%
\bibitem{Carm99} H.~J. Carmichael, \textit{Statistical Methods in 
  Quantum Optics 1: Master Equations and Fokker-Planck Equations} 
  (Springer-Verlag, Berlin, 2002).
%
\bibitem{Carmichael87} H.~J.~Carmichael, J. Opt. Soc. Am. B \textbf{4}, 
1588 (1987).
%
%
%
%
%
%
%
%




\end{references}
\end{document}